\documentclass[twocolumn]{aastex631}

\usepackage{pifont}
\usepackage{hyperref}
\usepackage{natbib}
\usepackage{amsmath}
\usepackage{amssymb}
\usepackage{eucal}
\usepackage{graphicx}
\usepackage{tikz}
\usepackage{xcolor}

\newcommand{\bagpipes}{\textsc{Bagpipes}}
\newcommand{\pylick}{\textsc{PyLick}}
\newcommand{\multinest}{\textsc{MuliNest}}
\newcommand{\cloudy}{\textsc{CLOUDY}}

\newcommand{\scipy}{\textsc{SciPy}}

\newcommand{\cmark}{\ensuremath{\cdot}}%
\newcommand{\xmark}{\ding{55}}%

\newcommand*{\figuretitle}[1]{%
    {\centering%
    \textbf{#1}%
    \par\medskip}%
}

\def\apjl{ApJL}
\def\apjs{ApJS}

\def\mailto#1{\href{mailto:#1}{#1}}

\received{May 6, 2021}
\revised{Februray 5, 2023}
\accepted{Februray 14, 2023}
\published{March 29, 2023}
\submitjournal{ApJS}
\shorttitle{Full-spectrum fitting of cosmic chronometers in the LEGA-C Survey}
\shortauthors{Jiao et al.}

\usepackage{soul} 

\begin{document}
\title{New Observational $H(z)$ Data from Full-Spectrum Fitting of Cosmic Chronometers in the LEGA-C Survey}

\author[0000-0003-0167-9345]{Kang Jiao}
\affiliation{Institute for Frontiers in Astronomy and Astrophysics, Beijing Normal University, Beijing 102206, China
; \mailto{tjzhang@bnu.edu.cn}}
\affiliation{Department of Astronomy, Beijing Normal University, Beijing 100875, China}
\affiliation{Dipartimento di Fisica e Astronomia ``Augusto Righi'', Alma Mater Studiorum Universit\`{a} di Bologna, via Piero Gobetti 93/2, I-40129 Bologna, Italy; \mailto{michele.moresco@unibo.it} }

\author[0000-0002-2889-8997]{Nicola Borghi}
\affiliation{Dipartimento di Fisica e Astronomia ``Augusto Righi'', Alma Mater Studiorum Universit\`{a} di Bologna, via Piero Gobetti 93/2, I-40129 Bologna, Italy; \mailto{michele.moresco@unibo.it} }
\affiliation{INAF - Osservatorio di Astrofisica e Scienza dello Spazio di Bologna, via Piero Gobetti 93/3, I-40129 Bologna, Italy}

\author[0000-0002-7616-7136]{Michele Moresco}
\affiliation{Dipartimento di Fisica e Astronomia ``Augusto Righi'', Alma Mater Studiorum Universit\`{a} di Bologna, via Piero Gobetti 93/2, I-40129 Bologna, Italy; \mailto{michele.moresco@unibo.it} }
\affiliation{INAF - Osservatorio di Astrofisica e Scienza dello Spazio di Bologna, via Piero Gobetti 93/3, I-40129 Bologna, Italy}

\author[0000-0002-3363-9965]{Tong-Jie Zhang}
\affiliation{Institute for Frontiers in Astronomy and Astrophysics, Beijing Normal University, Beijing 102206, China
; \mailto{tjzhang@bnu.edu.cn}}
\affiliation{Department of Astronomy, Beijing Normal University, Beijing 100875, China}
\affiliation{Institute for Astronomical Science, Dezhou University, Dezhou 253023, China}

\begin{abstract}
 In this work, we perform a full-spectrum fitting of 350 massive and passive galaxies selected as cosmic chronometers from the LEGA-C ESO public survey to derive their stellar ages, metallicities, and star-formation histories. We extensively test our results by assessing their dependence on the possible contribution of dust, calibration of noise and signal, and the use of photometric data in addition to spectral information; we as well identify indicators of the correct convergence of the results, including the shape of the posterior distributions, the analysis of specific spectral features, and the correct reproduction of the observed spectrum. We derive a clear age-redshift trend compatible with the aging in a standard cosmological model, showing a clear downsizing pattern, with more massive galaxies being formed at higher redshift ($z_f\sim2.5$) with respect to lower massive ones ($z_f\sim2$). From these data, we measure the differential aging of this population of cosmic chronometers to derive a new measurement of the Hubble parameter, obtaining  $H(z=0.8)=113.1 \pm15.1(\mathrm{stat.})^{+29.1}_{-11.3}(\mathrm{syst.})$. This analysis allows us for the first time to compare the differential ages of cosmic chronometers measured on the same sample with two completely different methods, the full-spectrum fit (this work) and the analysis of Lick indices, known to correlate with the age and metallicity of the stellar populations \citep{Borghi2022a}. Albeit an understood offset in the absolute ages, the differential ages have proven to be extremely compatible between the two methods, despite the very different data, assumptions, and models considered, demonstrating the robustness of the method.
\end{abstract}


\section{Introduction} \label{sec:introduction}

    Since the discovery of the accelerating expansion of the Universe \citep{Riess1998,Perlmutter1999}, the cosmological community has been working to understand the mechanism of this expansion. Modern cosmology postulates that dark energy, an unknown form of energy with negative pressure, is driving the accelerated expansion of the late Universe and that the gravitational effect of Cold Dark Matter (CDM) shapes the large-scale structure of the Universe, a model dubbed $\Lambda$CDM. Numerous cosmological probes and observations, including the Cosmic Microwave Background (CMB, eg. \citealt{Smoot1992ApJ,Bennett2003ApJS,Planck2014AA,Swetz2011ApJS,Carlstrom2011PASP,Planck2020AA}), Baryon Acoustic Oscillations (BAO, eg. \citealt{Percival2001MNRAS.327.1297P,Cole2005MNRAS.362..505C,Eisenstein2005ApJ...633..560E}), Type Ia supernovae (SNe, eg. \citealt{Sullivan2011ApJ,Suzuki2012ApJ,Betoule2014AA,Scolnic2018ApJ}), weak gravitational lensing \citep{Bartelmann2001PhR...340..291B}, and cluster counts \citep{Allen2011ARA&A..49..409A}, have been proposed and extensively studied to determine the Universe's large-scale structure and evolution. After more than twenty years of unremitting efforts, we are now in the golden age of precision cosmology, with measurements and constraints on cosmological parameters reaching the percent level.

    The Hubble constant $H_0$ has long been a critical observable of observational cosmology \citep{freedmanHubbleConstant2010}, and its value is directly related to our current estimate of the Universe's age. However, the two probes that represent the most precise level of measurement today, SNe and CMB, produce significant discrepancies beyond the order of $4\sigma$ (for an extensive review, see \citealt{Valentino2021CQGra}). The increase of observational evidence supporting this discrepancy between observations of the early and late Universe, has undoubtedly kicked off a crisis in modern cosmology \citep{Verde2019,davisExpandingControversy2019,Riess2020,Abdalla2022}. At the moment, there are suggested theories to try to explain it, even if not definitive \citep{Valentino2021CQGra}. Alternative cosmological probes \citep{Moresco2022} can play an important role in obtaining additional independent, high-precision measurements to assess the current Hubble tension's reliability. It also becomes evident that a single probe is not adequate to constrain the properties and evolution of the Universe accurately and completely. The Hubble parameter $H(z)$ is the physical quantity that most directly describes the history of the Universe's expansion, and its measurement has advanced significantly over the last decade or so. We can not only reconstruct these $H(z)$ measurements to extrapolate $H_0$ at zero-redshift, shedding light on a different path to explore the crisis, but also enhance our ability to understand the nature of dark energy, which dominates the late Universe just covered by the observable range of $H(z)$. 

    Existing observational $H(z)$ data (referred to as OHD, see e.g. \citealt{Zhang2010AdAst,Ma2011ApJ}) are mainly based on two probes: the differential age method and the radial BAO size method \citep{benitez2009}. The former can be obtained with Cosmic Chronometers (CC, \citealt{Jimenez2002ApJ,Moresco2018ApJ,Moresco2020ApJ}) by measuring the differential age-redshift relation of massive and passive galaxies throughout the Universe. Any systematic offset introduced by the galaxy age measurement method will be canceled out when deriving the differential age. A total of 32 H(z) measurements have been obtained \citep{Jimenez:2003,Simon:2005,Stern:2010,Moresco:2012,Zhang:2014,Moresco:2015,Moresco:2016,Ratsimbazafy:2017,Borghi2022b} and are currently widely used to test cosmological models. These measurements are regarded as cosmological model-independent since the principle is not dependent on the choice of cosmological models. The second method is based on the inverse proportionality between H(z) and the differential radial (comoving) distance, which can be traced by measuring the radial size of BAO features at different redshifts. This method, however, requires knowledge of the comoving BAO scale ($r_\mathrm{BAO}$), which is derived from the CMB measurements. This fact makes this probe not fully cosmology-independent, since typically in the derivation of the sound horizon scale from CMB a cosmology model is assumed. Additionally, gravitational waves can be used as standard sirens \citep{Schutz1986Nature,Holz2005ApJ,Abbott2017Natur}, to study H(z), with promising perspectives for the next decade \citep{Farr2019}. Finally, the phenomenal growth of Fast Radio Burst (FRB) observations also expands the $H(z)$ measurement possibilities \citep{wuNewMethodMeasure2020}.

    Selecting a pure passive sample and measuring the age difference between galaxies are the two bases of the CC method. Various strategies have been proposed to distinguish `passive' from `star-forming' galaxies, including morphological selections of spheroidal systems (following \citealt{1936rene.book.....H}), cuts on color-color diagrams (e.g. UVJ, \citealt{Williams2009ApJ}; NUVrJ, \citealt{Ilbert2013}) or on a color-mass diagram (e.g. \citealt{Peng2010ApJ}), and Spectral Energy Distribution (SED) fitting (e.g. \citealt{Pozzetti2010}). Combining multiple criteria and maximizing the overlap of complementary information (photometric and spectroscopic) result in a significantly more effective method of selecting a pure sample \citep{Moresco2013, Moresco2018ApJ,Borghi2022a}.  While spectral line analysis enables us to obtain extremely precise redshift values, the situation is much more complicated in determining the age, which can not be directly observed, but can be estimated using photometry (SED), single spectral regions (e.g., D4000, Lick indices), or the entire spectrum features (full-spectral fitting).
    However, each of these methods may suffer various systematics caused by parameter degeneracies. \cite{morescoEARLYTYPEGALAXIESPROBES2011} explains that SED-fitting, which is commonly used to derive galaxies' ages, is incapable of fully breaking the age-metallicity degeneracy; also, the age degenerates to $\tau$ in the delayed exponential Star-Forming History (SFHs). \cite{Moresco2011jcap} proposed an innovative method that consists in not using the age but rather a direct spectroscopic observable (the 4000~{\AA} break) to measure $H(z)$, making the decoupling of systematic and statistical error easier.  Most recently, \cite{Borghi2022b} obtained for the first time a new $H(z)$ measurement using the Lick indices method. Their analysis takes advantage of the high signal-to-noise (S/N) spectroscopic data of the LEGA-C (DR2) survey \citep{Straatman2018} of galaxies at $z\sim0.7$. However, the Lick indices method does not allow to flexibly study the galaxies' star formation histories, which are useful to better exclude possible biases in the derived age-redshift relation and, therefore, on H(z). For the purpose of optimizing the set of the spectral absorption features for Lick indices fitting, \cite{Borghi2022b} only make use of a sub-sample (140 galaxies) of the selected passive sample, while the full-spectrum fitting is not subject to this issue. 

    In this paper, we perform full-spectrum fitting to derive the ages and star-formation histories of passive galaxies in LEGA-C DR2, then use them as cosmic chronometers to obtain a new $H(z)$ measurement. The dataset is introduced in Section~\ref{sec:data}, and the fundamental principles and details of the full-spectrum fitting in Section~\ref{sec:method}. In Section~\ref{sec:analysis} we present the results on physical parameters and the strategies to improve the performance of their estimation. In Section~\ref{sec:hubble_parameter} we detail the procedure for applying the CC method, presenting and discussing the final $H(z)$ measurement results. The conclusions are presented in Section~\ref{sec:conclusions}.


\section{Data} \label{sec:data}

    In this section, we describe the spectroscopic and photometric data used in this analysis, and the selection criteria adopted to select the sample of cosmic chronometers.

    \begin{figure*}[t]
        \includegraphics[width=.95\hsize]{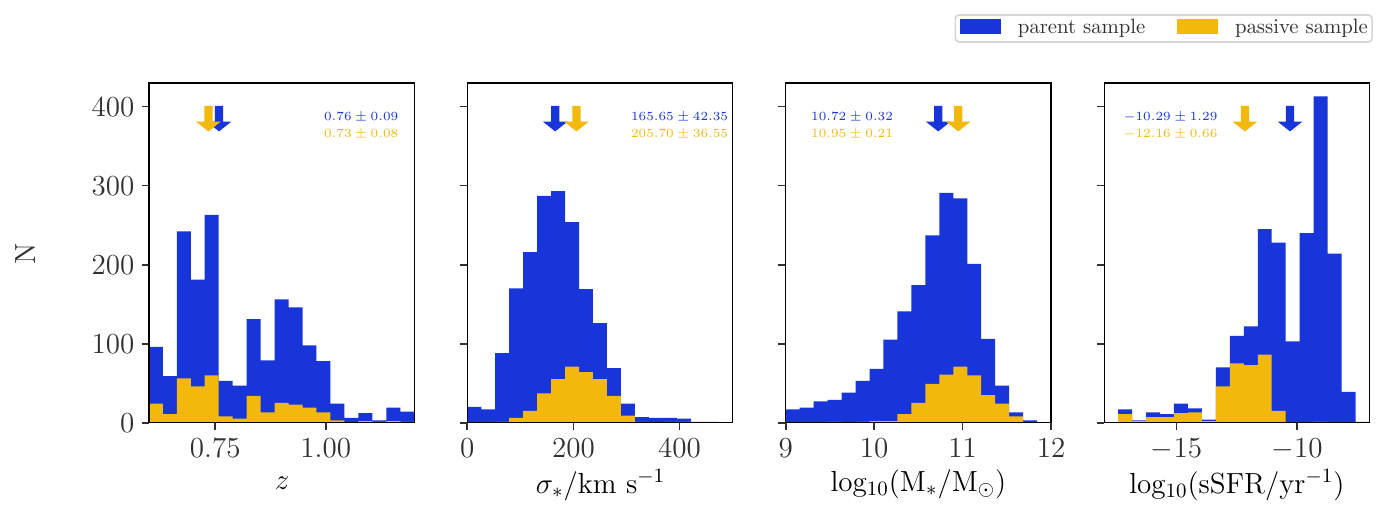}
        \caption{Distributions of four key parameters for the LEGA-C DR2 parent sample (blue) and the 350 passive galaxies analyzed in this work (yellow). The redshift ($z$), stellar velocity dispersion ($\sigma_\star$) and stellar mass ($M_\star$) are taken from LEGA-C DR2, while the specific star formation rate (sSFR) are from COSMOS2015. The arrows mark out the median values.}\label{fig:hist_sample}
    \end{figure*}

    \paragraph{\bf Spectroscopic Data}
    The spectroscopic data are taken from the Large Early Galaxy Astrophysics Census (\mbox{LEGA-C}), an ESO 130-night public survey of $\sim3200$ $K_s$-band selected galaxies conducted with VLT/VIMOS \citep{lefevre2003SPIE.4841.1670L} on the Very Large Telescope. The 20-hour long integrations produce continuum spectra with an average  $S/N \sim 21.8 $ per pixel ($0.6$~{\AA}) for massive galaxies ($M\gtrsim10^{11}M_\odot$). The second data release (LEGA-C DR2, \citealt{Straatman2018}) includes 1988 spectra in the redshift range $0.6\lesssim z \lesssim 1.0$ covering the observed wavelength range $\sim 6300-8800$~{\AA}, with an effective spectral resolution of $R \sim 3500$. We add to the LEGA-C dataset the spectral indices measurements from \citet{Borghi2022a}, providing a catalog of Lick indices measurement including also the recent CaII H/K diagnostic \citep[a useful tracer of recent episodes of star formation,][]{Moresco2018ApJ}.

    \paragraph{\bf Photometric Data} 
    One of the advantage of the LEGA-C sources is that, being observed in the COSMOS field, a wealth of multi-wavelength photometric observations are available (e.g.,  \citealt{Muzzin2013ApJS,Laigle2016ApJS,Weaver2022ApJS}). In this work, following \cite{Straatman2018} we adopt the Ultra Deep Survey with the VISTA telescope (UltraVISTA) photometric catalog from \citep{Muzzin2013ApJS}. We use a total of 21 photometric bands, namely, IB427, IB464, IA484, IB505, IA527, IB574, IA624, IA679, IB709, IA738, IA767, IB827, u, V, zp, Y, J, H, Ks, ch1, ch2. For a given filter $x$, we compute the total flux $f_{x, \mathrm{tot}}$ by applying the equation 
    \begin{equation}
    f_{x, \mathrm{tot}}=f_{x} \times \frac{f_{K_s, \mathrm{tot}}}{f_{K_s}},
    \end{equation}
    where $f_{K_s, \mathrm{tot}}$  is the total $K_s$ band flux from SExtractor's \textit{flux\_auto} that has been corrected using the growth curve of the point-spread function (PSF) stars and $f_{K_s}$ is the specific Ks-band flux  \citep[see][]{Muzzin2013ApJS}. The SEDs of the catalogues are in good agreement, but differences in calibration and measurement precision may affect the age-$z$ relation. We will further explore the use of different photometric calalogues in a follow-up analysis.

    \paragraph{\bf The Sample}\label{par:bonafide_passive_sample}

    By combining NUVrJ selection, a cut based on the equivalent width of the [OII]$\lambda$3727 line $< 5$~{\AA}, and a visual inspection to further remove galaxies with strong [OII]$\lambda$3727 and/or [OIII]$\lambda$5007 emission lines, \cite{Borghi2022a} selected a pure sample of 350 passive galaxies in LEGA-C DR2, minimizing any residual contamination from star-forming outliers. The distribution of some key parameters describing this population is shown in Figure~\ref{fig:hist_sample}. This passive sample has a median redshift of $\left\langle z\right\rangle=0.735$ with two peaks around $z\sim0.745$ and $z\sim0.839$. 
    The median values of the $\sigma_\star$ and $\log_{10}(M_\star/M_\odot)$ distributions increases from $165.7\ \mathrm{km\ s^{-1}}$ ($10.72$) to $205.7\ \mathrm{km\ s^{-1}}$ ($10.95$) respectively.
    Its specific star formation rate (sSFR) distribution has a median logarithmic value of $\left\langle\mathrm{log_{10}(sSFR)}\right\rangle=-12.2\ \mathrm{{yr}^{-1}}$, which is $\sim1$~dex lower than what is typically used to define a galaxy `passive' \citep[e.g.,][]{Pozzetti2010AA}. 

    \begin{figure*}
        \centering
            \includegraphics[scale=.6]{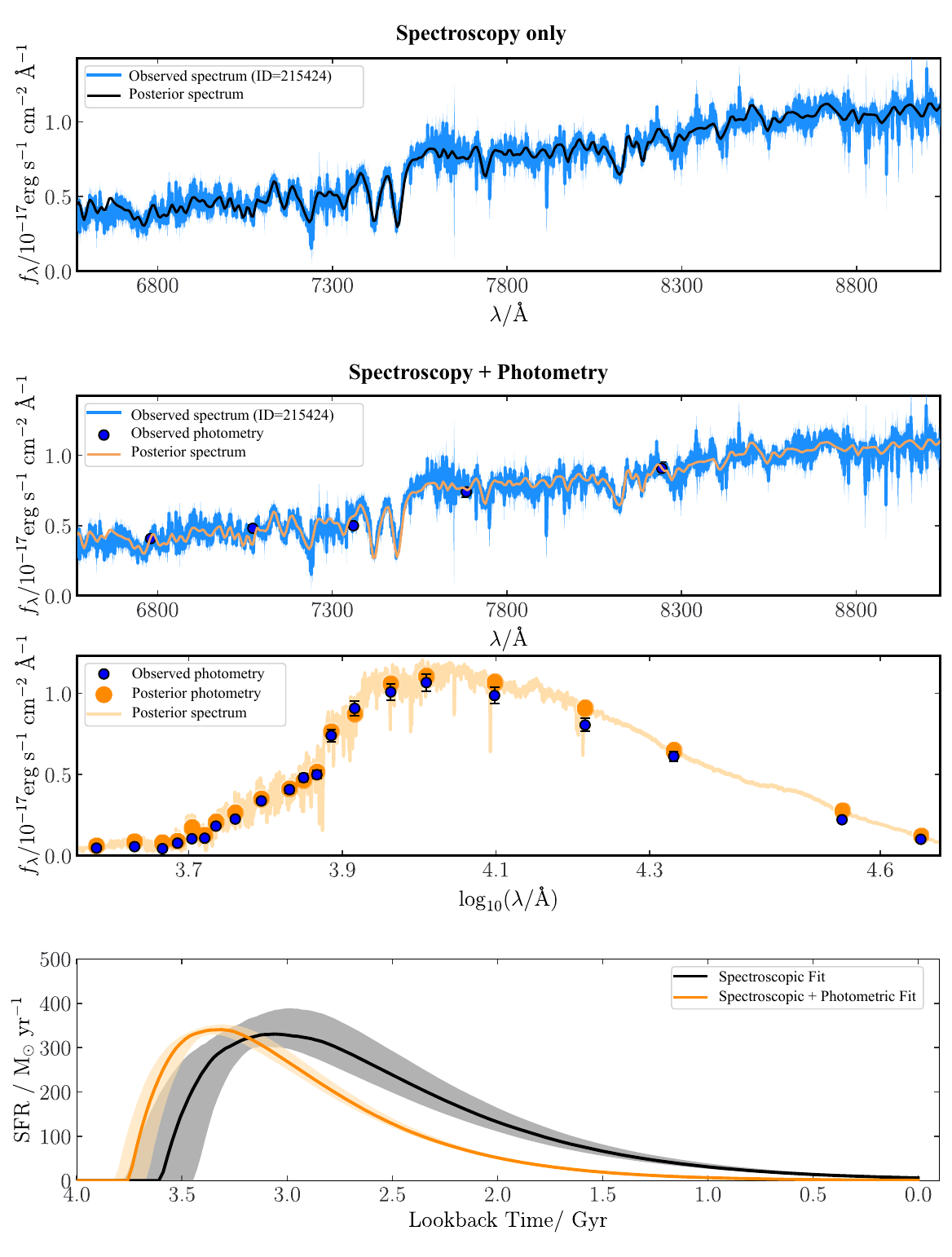}
            \caption{Full spectrum fitting results for an example galaxy (ID$=$215424) obtained with spectroscopic data alone (top panel) and adding photometry (second and third panel). The observational spectra from LEGA-C DR2 \citep{Straatman2018} and photometric data points from the UltraVISTA catalogue of \cite{Muzzin2013ApJS} are shown in light blue and blue, respectively. The best-fit BAGPIPES spectra and photometric points from fitting spectroscopy and spectroscopy$+$photometry are shown in black and orange, respectively. The bottom panel shows the best-fit BAGPIPES SFR (as a function of lookback time) correspondingly, the solid curves are the median posteriors while the shade regions are the $1\sigma$ confidence regions, the horizontal axis is the lookback time since $t(z_\mathrm{obs})$. }\label{fig:spectrum&SFH2}
    \end{figure*}

\section{Method}\label{sec:method}
    
    \paragraph{\bf Full Spectrum Fitting}
    To perform the full-spectrum fitting we use the \bagpipes\ code developed by \citet{Carnall2018MNRAS}. \bagpipes\ models the observed spectrum  of a galaxy $f_{\mathrm{obs}}$ into $f^{\mathcal{H}}(\Theta)$ based on an hypothesis $\mathcal{H}$ of the physics involved described by parameters $\Theta$. The posterior distribution $P(\Theta \mid f_\lambda , \mathcal{H})$ obtained from the Bayes theorem
    \begin{equation}
        P(\Theta \mid f_{\lambda}, \mathcal{H})=\frac{\mathcal{L}(f_{\lambda} \mid \Theta, \mathcal{H}) P(\Theta \mid \mathcal{H})}{P(f_{\lambda} \mid \mathcal{H})},
    \end{equation}
    is sampled with the nested sampling algorithm \multinest\ \citep{2014BuchnerMULTINEST}. Here $\mathcal{H}$ includes the modelling of the star formation rate $\mathrm{SFR}(t_{\mathrm{U}i})$, the simple stellar-population $\mathrm{SSP}(t_i, \lambda, Z_j)$, and the neutral and ionized interstellar medium (ISM) radiative transmission function $T^0(t_i, \lambda)$ and $T^+(t_i, \lambda)$, which are used to simulate the luminosity function of a galaxy,
    \begin{equation}
    \label{pipes_model_equation}
    \begin{split}
    L_\lambda = \sum_{j=1}^{N_{c}} \sum_{i=1}^{N_a} \mathrm{SFR}_j&(t_{\mathrm{U}i})\ \mathrm{SSP}(t_{\mathrm{U}i}, \lambda, Z_j)\times \\  &\times T^+(t_i, \lambda)\ T^0(t_i, \lambda)\ \Delta t_i, 
    \end{split}
    \end{equation}
    where $t_\mathrm{U}, t, \lambda, Z$ are the cosmic time, the age of the stellar population, the wavelength of spectral line and stellar metallicity, respectively, and the subscripts $i$ and $j$ denote summations for all the age bins and SFH components, respectively. $\mathcal{H}$ also includes the modelling of the intergalactic medium (IGM) radiative transfer to finally simulate the observed flux,
    \begin{equation}
    \label{eq:flux}
        f_{\lambda_{\mathrm{obs}}}=\frac{L_{\lambda}}{4 \pi D_{L}(z_{\mathrm{obs}})^{2}\left(1+z_{\mathrm{obs}}\right)} T_{\mathrm{IGM}}\left(\lambda, z_{\mathrm{obs}}\right),
    \end{equation} where $\lambda_{\mathrm{obs}} = (1+z_{\mathrm{obs}})\lambda$, $D_L(z_{\mathrm{obs}})$ is the luminosity distance and $T_{\mathrm{IGM}}$ is the transmission function of the IGM. The nebular emission lines and continuum come from pre-computed \cloudy~\citep{Ferland2017RMxAA} grids with only one free parameter, the logarithmic ionization parameter ($\log_{10}(U)$). We apply the \cite{Charlot2000ApJ} model out of the four choices of dust attenuation models (see detail descriptions in \citealt{Carnall2018MNRAS}) that \bagpipes~provides.  
    The likelihood function can be written in a logarithmic form as,
    \begin{equation}
        \ln (\mathcal{L})=-0.5 \sum_{i} \ln \left(2 \pi \sigma_{i}^{2}\right)-0.5 \sum_{i} \frac{\left(f_{i}-f_{i}^{\mathcal{H}}(\boldsymbol{\Theta})\right)^{2}}{\sigma_{i}^{2}},
    \end{equation}
    where $\sigma$ is the observation error of the fluxes, and here sums over all $i$-th wavelength pixels.

    In addition to the spectrum, \bagpipes~allows the inclusion of photometric data points in the fit, thus enabling modelling a galaxy SED on a wide wavelength range, from far-ultraviolet to microwave regimes. Another significant advantage is that it is possible to adaptively test different SFH choices (e.g., single burst, constant, exponentially declining, as well as a combination of them). Models within the \bagpipes\ code are resampled in an age grid $\Delta t_i$ based on the SSP model generated using the 2016 version of the \cite{BC2003MNRAS} (BC16) models. \bagpipes~ is structured around three core classes, which are \texttt{galaxy} for loading observational data, \texttt{model\_galaxy} for generating model galaxy spectra and \texttt{fit} for fitting models to observational data.  The code is open source and publicly avaliable\footnote{\url{https://github.com/ACCarnall/bagpipes}}.
    To extract parameters values and associated uncertainties from the posterior distributions, we adopt the median and 16--84$^\mathrm{th}$ percentiles, respectively.

    \paragraph{\bf Star Forming History Choice} 
    Most of the CC analyses, including \cite{Borghi2022b}, assume single burst star formation histories as an ideal simplification of the real SFH. This model assumes that the total mass of a galaxy suddenly formed at a specific cosmic time, which is characterized by a delta function $\mathrm{SFR}(t_\mathrm{U})\propto\delta(t_\mathrm{U})$. A more realistic SFH model is necessary to test the robustness of the age-$z$ relations and the $H(z)$ obtained. In this work, we extend this analysis by testing two other well-established SFH models, namely the double power law (DPL) and the delayed exponentially declining (DED) model based on the CC sample that \cite{Borghi2022a} compiled. The DPL model separates the rising and declining phases of the SFH using two separate power-law slopes, $\mathrm{SFR}(t_\mathrm{U}) \propto \left[ (t_\mathrm{U}/\tau)^{\alpha} + t_\mathrm{U}/\tau)^{-\beta}\right]^{-1}$, where $\alpha$ is the falling slope, $\beta$ is the rising slope and $\tau$ is related to (but not the same as) the peak time. 
    The DED model assumes that the star-formation starts at some time $T_0$ and increases gradually to its peak, after which it declines exponentially with some timescale $\tau$,
    \begin{equation}
        \operatorname{SFR}\left(t_{\mathrm{U}}\right) \propto \begin{cases}
        \left(t_{\mathrm{U}}-T_{0}\right) \mathrm{exp}\left(-\frac{t_{\mathrm{U}}-T_{0}}{\tau}\right) & t_{\mathrm{U}}>T_{0} \\ 0 & t_{\mathrm{U}}<T_{0}
        \end{cases}.
        \label{eq:ded}
    \end{equation}
    As detailed in this Section~\ref{sec:method}, we fit the selected sample separately using the DPL and DED model and get compatible median reduced chi-squared values for the spectrum of $\chi^2_\nu=1.96$ and $\chi^2_\nu=1.92$. According to the principle of the Ockham's Razor, we choose the one fewer free parameter model - DED (Equation~\ref{eq:ded}) for our following analysis.

    \paragraph{\bf Removing Cosmological Prior} 
    The age of a stellar population, $t$, is defined as the look-back time between its observed redshift and the beginning of its star-formation,
    \begin{equation}
        t \equiv t_\mathrm{U}(z_\mathrm{obs}) - t_\mathrm{U,\,start}.
    \end{equation}
    In fact, the cosmic time ($t_\mathrm{U}$) at a given redshift is not a direct observable, we can only calculate its value based on a cosmological model. \bagpipes~use $\Lambda$CDM as its default cosmological model, with the default parameters $\Omega_M = 0.3$, $\Omega_\Lambda = 0.7$ and $H_0$ = 70 $\mathrm{km\ s^{-1}\ Mpc^{-1}}$.  In principle, when the age of a galaxy exceeds the age of the Universe, it is reasonable to consider it non-physical. To address this issue, \bagpipes\ 
    assumes that the star formation rate $\mathrm{SFR}(t_{\mathrm{U}i})=0$ when the retrieved age is larger than the estimated age of the Universe at the given redshift. 
    While this assumption is typically neglected in galaxy evolution studies, being of relative interest for the results, imposing an upper limit on the retrieved age based on a cosmological model is to be strictly avoided in our analysis. In particular, such a prior could induce cosmological biases in the age estimates and circular arguments in the derivation of the Hubble parameter, since if the ages of the oldest objects are set to the age of the Universe of the reference cosmological model the method would artificially provide, by definition, the reference cosmology.

    Fortunately, we can avoid the above situation by releasing the upper limit for $t_U$ in \bagpipes\ to a value that our sample galaxies cannot exceed, such as 20~Gyr at all redshifts. This modification changes the upper boundary of the age sampling, without affecting the sampling grid. 
    Galaxies' age can be easily retrieved by subtracting the formation time with the new upper limit we set, and the age-$z$ slop will not be affected by the cosmological assumptions. 
    Besides, we notice that the cosmological model is also used when calculating the $D_L$ in Equation~\ref{eq:flux}. The $D_L$ doesn't interact with $t_U$ in the rest of the code and because of the negligible dependence of $\mathrm{d}H(z)\sim\mathcal{O}^2(\mathrm{d} D_L)$, it is acceptable to ignore the issue of $D_L$ affected by the choice of cosmological model (see figure 1 in \citealt{Jimenez2002ApJ}). 
    In conclusion, with these modifications we ``erase''  the effect of the cosmological prior on the galaxy age estimation in the original \bagpipes\ code.

    \paragraph{\bf Adding Photometric Data}
    The spectroscopy covers a relatively narrow wavelength range of the galaxies' entire spectrum compared with the photometry. Fitting spectroscopic data alone, due to the lack of enough information, is incapable of fully modelling the line features and breaking the degeneracy between parameters, especially the age-$\tau$ degeneracy in our analysis, as well as the CaII H and K lines that are essential for diagnostic of passive galaxies. Adding photometric data will improve the performance of fitting by providing additional information. We employ the Ks selected UltraVISTA photometries in our analysis as detailed in Section~\ref{sec:data}. 

    In Figure~\ref{fig:spectrum&SFH2} we show the full spectrum fitting results from an example galaxy (ID~215424) obtained with spectroscopic data alone and adding photometry. We observe that in the posterior the CaII~K line is less deep than the H line for the spectroscopic (only) fitting, contradicting the observational data. On the contrary, this feature is well reproduced after including the photometric data in the fit. This same behavior is observed, in general, for the entire sample. In particular the median percentage difference between the observed and reconstructed $\mathrm{H/K}$ (see also Table~\ref{Table:Priors} and Section~\ref{sub:breaking_degeneracies} for a more extended discussion) is significantly reduced from the value of  $11.93\pm 6.76\ (\%)$ to $6.46 \pm4.34\ (\%)$. 
    From the histograms in Figure~\ref{fig:hist_posteriors}, we observe both long tails of derived ages and $\tau$ distributions for the spectroscopic (only) fitting, indicating the existence of the age-$\tau$ degeneracy, while the tails are significantly suppressed after adding the photometric fitting.

    \begin{deluxetable*}{cccccccccccc}
        \tablewidth{0em}
        \tabletypesize{\footnotesize}
        \tablecaption{ Model ingredients for the various analyses performed.}\label{Table:Priors}
        \tablehead{\colhead{$z$} & \colhead{$t$/Gyr} & \colhead{$\tau$/Gyr} & \colhead{$Z/Z_\odot$} & \colhead{$\log(M_\mathrm{form}/M_\odot)$} & \colhead{$\sigma_\star/\mathrm{km\,s^{-1}}$} & 
        \colhead{calib.} & \colhead{noise} & \colhead{dust} & \colhead{photo.} & \colhead{$\left\langle\chi^2_\nu\right\rangle$} & \colhead{$\left\langle\Delta\:\mathrm{H/K}\right\rangle$} \\ 
        \colhead{\textit{fixed}} & \colhead{$\mathcal{U}[0,20]$} & \colhead{$\mathcal{U}[0,2]$} & \colhead{$\mathcal{U}[0.001,3]$} & \colhead{$\mathcal{U}[0, 18]$} & \colhead{$\mathcal{G}[\sigma_\star;err_{\sigma_\star}]$} & 
        \colhead{} & \colhead{} & \colhead{} & \colhead{21 bands} & \colhead{} & \colhead{\%}} 
        \colnumbers
        \startdata
        \cmark & \cmark & \cmark & \cmark & \cmark & \cmark & \xmark & \cmark & \cmark & \xmark & $1.95\pm0.44$ & $11.93\pm6.76$\\
        \cmark & \cmark & \cmark & \cmark & \cmark & \cmark & \cmark & \cmark & \xmark & \cmark & $1.96\pm0.45$ & $9.66\pm5.94$\\
        \cmark & \cmark & \cmark & \cmark & \cmark & \cmark & \cmark & \cmark & \cmark & \cmark & $1.93\pm0.43$ & $7.99\pm5.34$ \\
        \cmark & \cmark & \cmark & \cmark & \cmark & \xmark & \cmark & \cmark & \cmark & \cmark & $1.93\pm0.43$ & $7.83\pm5.17$ \\
        \cmark & \cmark & \cmark & \cmark & \cmark & \cmark & \cmark & \xmark & \cmark & \cmark & $1.93\pm0.43$  & $6.58\pm4.08$\\
        \hline
        \cmark & \cmark & \cmark & \cmark & \cmark & \cmark & \xmark & \cmark & \cmark & \cmark & $1.97\pm0.45$ & $6.46\pm4.34$ \\
        \enddata
        \flushleft{
        \tablecomments{The \cmark \ and \xmark \ mark the use or not of the corresponding prior (Columns 1--6, see Section~\ref{sec:method}), the model components (Columns 7--9; three-order polynomial calibration, white noise, and the \citealt{Calzetti2000ApJ} dust attenuation model), and the photometric data (Column 10; see Section~\ref{sec:data}). The $\left\langle\chi^2_\nu\right\rangle$ and $\left\langle\Delta\:\mathrm{H/K}\right\rangle$ are the median $\pm$ median absolute deviation of the distributions taking account of the 350 fitting results.}
        }
    \end{deluxetable*}

    \paragraph{\bf Exploring Parameter Space}
    In our analysis, we fully explore all the possible parameter space consist of the physical properties of the galaxy, the parameters of modelling the noise and the dust attenuation, together with the calibration of the fluxes. We consider two diagnostics, the reduced chi-square ($\chi^2_\nu$) and the percentage difference of $\mathrm{H/K}$ (see topical description in Section~\ref{sub:breaking_degeneracies}) to quantify the agreement between the posterior and observed spectrum and to verify the improvement of the fit when adding more information on the parameters.  

    Fitting redshift is not necessary, for each galaxy we set its $z$ to the value of the high precision spectroscopic redshift derived from numerous high S/N absorption features observed the LEGA-C DR2 spectra. We take Uniform priors in reasonably wide parameter spaces for the age $(t)$, star formation timescale ($\tau$), stellar metallicity $(Z)$, and the $\log$ of the mass formed $\log_{10}(M_\mathrm{form})$ as described in Table~\ref{Table:Priors}. The stellar velocity dispersion $\sigma_\star$ is a direct observable derived from the line-broadening of the spectrum, we test the effect of using a wide uninformative prior $\sigma_\star\sim\mathcal{U}[0,400]$ and narrower Gaussian prior of $\sigma_\star\sim\mathcal{G}[\sigma_\star;err_{\sigma_\star}]$, set by the measurements provided in the LEGA-C DR2 catalog. We also test the impact of varying the observed flux calibration according to a second-order Chebyshev polynomial with coefficients' prior set to $P_0 \sim \mathcal{G}[1, 0.25]$, $P_1 \sim\mathcal{G}[0, 0.25]$ and $P_2 \sim\mathcal{G}[0, 0.25]$ for each of the orders respectively. We also test the white noise model as an additional component, by adopting a uniform prior for the logarithmic white noise scaling parameter $\log_{10}(S_{noise})\sim\mathcal{U}[0,10]$. Finally, even if we expect it to be negligible for this sample, we additionally model the dust effect by assuming the \cite{Calzetti2000ApJ} attenuation curve and a Gaussian prior on the absolute attenuation at $5500$~{\AA}, $A_V\sim\mathcal{G}[1, 0.25]$, a multiplicative factor of $\eta=2$ for stars in birth clouds, and an attenuation power-law slope of $n\sim\mathcal{G}[0.7, 0.3]$ in the range $n\in[0.3, 2.5]$. For further details on each model component please check \cite{Carnall2018MNRAS}.  When doing the analyses, we do not expect the nebular and dust emission modellings to have significant impacts on our results, since our galaxies were selected to have negligible or no emission lines contribution \citep{Borghi2022b}.

    We compare the results obtained from using or not the aforementioned components and find all the $\left\langle\chi^2_\nu\right\rangle$ are compatible (on the second digits).  The modelling of $\mathrm{H/K}$ (see topical description in Section~\ref{sub:breaking_degeneracies}) are significantly improved (on the first digits of $\left\langle\Delta\:\mathrm{H/K}\right\rangle$) when not using the calibration or modelling the noise. Since the LEGA-C DR2 spectra are flux-calibrated using UltraVISTA's photometric SEDs, we prefer not to impact on the data with an additional calibration. But the other option is also acceptable and we get compatible $H(z)$ measurements for them during our following step analyses. We decide to use as a baseline the set of parameters that better reproduces the CaII H/K feature (the last row of Table~\ref{Table:Priors}, since it has been proven to be a very powerful and important diagnostic to trace the purity of CC samples \citep[see][]{Moresco2018ApJ,Borghi2022a,Moresco2022} and having a significantly different H/K in the reconstructed spectrum would mean not to correctly reproduce the behaviour of our data, possibly resulting in biases in the results; moreover, we notice that all the models present a compatible $\chi^{2}_{\nu}$, but this model has also the advantage to avoid additional calibration, since the LEGA-C DR2 spectra are flux-calibrated using UltraVISTA's photometric data.

\section{Analysis} \label{sec:analysis}

    In this section, we present our analysis of the physical parameters derived. We start by defining three diagnostic criteria to assess the reliability of our results, and flag the constraints not properly converged. We will explore their use to improve the robustness of the derived parameter, if needed, concluding by presenting our baseline results on which will be based the cosmological analysis.

    \subsection{Breaking The Degeneracies}\label{sub:breaking_degeneracies}

\begin{figure*}[htb!]
        \centering
        \vspace{1.5em}\figuretitle{Spectroscopic (only) Fit}
        \includegraphics[trim={1cm 0 0 0}, width=.95\hsize]{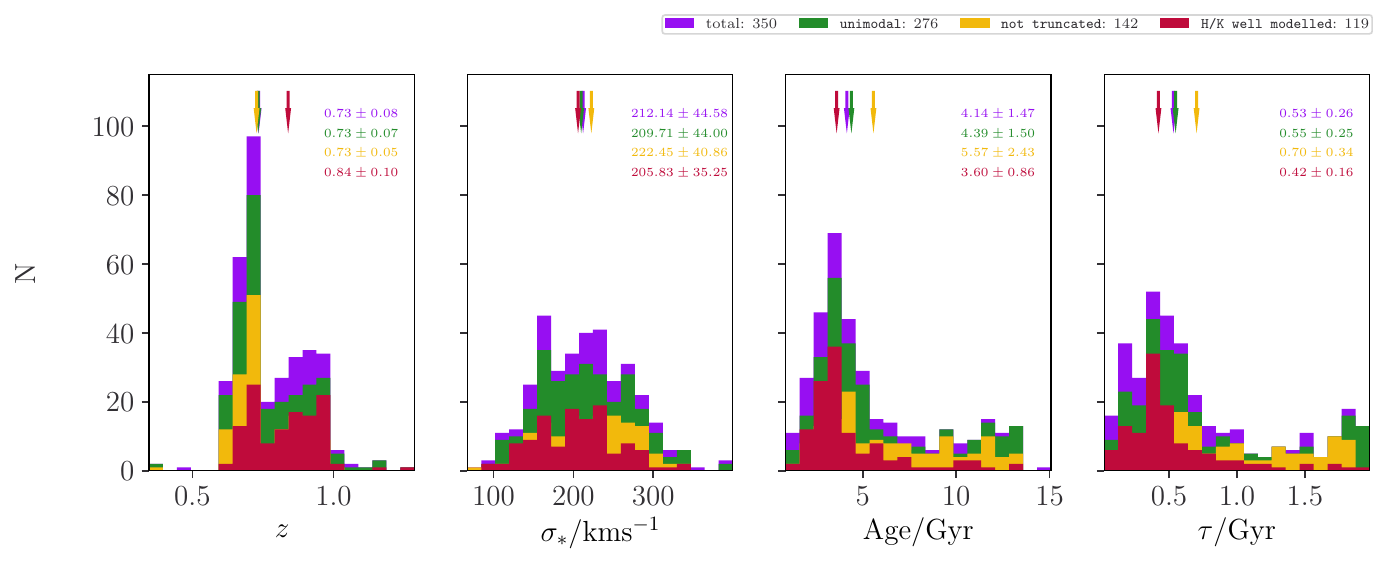}
        \\\vspace{1.5em}\figuretitle{Spectroscopic + Photometric Fit}
        \includegraphics[trim={1cm 0 0 0},width=.95\hsize]{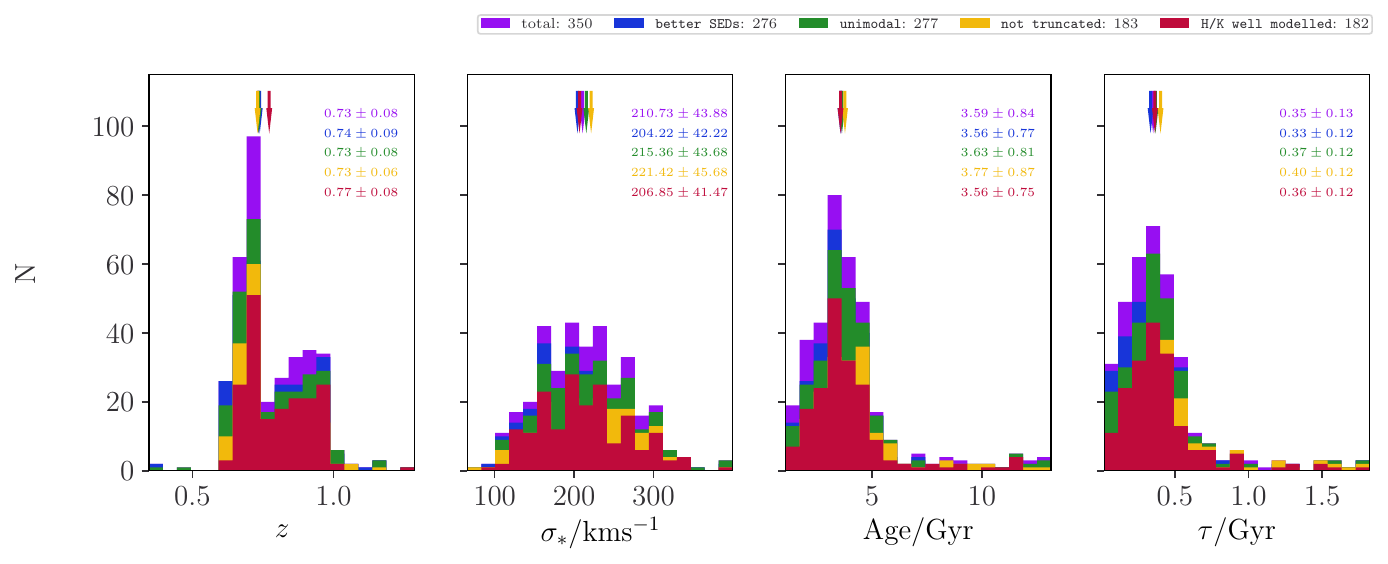}
        \caption{Distributions of redshift ($z$), stellar velocity dispersion ($\sigma_\star$), age, and formation time scale ($\tau$) obtained from the fit of LEGA-C DR2 cosmic chronometers with spectroscopic data alone (upper panels) and adding photometric data from \citet{Muzzin2013ApJS} (lower panels). In each panel, we show the histograms of the full sample (purple) and those obtained when applying the quality flags described in Section~\ref{sub:breaking_degeneracies}, namely, \texttt{better SEDs} (blue), \texttt{unimodal} (green), \texttt{not truncated} (yellow), \texttt{H/K well modelled} (red).}\label{fig:hist_posteriors}
    \end{figure*}

    In Figure~\ref{fig:hist_posteriors} we present the distributions of the derived parameters ($z$, $\sigma_\star$, stellar age, and $\tau$) obtained from our analysis of both the spectroscopic data alone and from the fit of the spectroscopic and photometric data combined. We start by observing that the pure spectroscopic case presents significant tails in the age and $\tau$ distributions, with ages larger than the age of the Universe and $\tau$ up to the maximum value allowed by the prior. We will therefore define a set of diagnostic criteria to check the accuracy of our results, apply them to our constraints, and verify if and how they can impact the nonphysical large values just discussed.

    \paragraph{\bf Improperly Converged Constraints} 
    The values estimated from the posterior distributions obtained from \bagpipes~ may not be fully reliable depending on several issues. As an example, this is the case when the posterior distribution exhibits multi-modal peaks, or is very skewed toward the edge of the domain allowed by the priors. Considering the large number of galaxies and combinations of parameters explored, we develop two efficient automatic algorithms to assist us in recognizing these issues and subsequently flagging the corresponding galaxies.
    
    We identify the multi-modality by counting the convex inflection points of the one-dimensional posterior distribution functions (PDFs) after applying a Gaussian kernel smoothing function with a bandwidth with size 12\% of the full posterior range. This technique is robust against spurious detection of close-by peaks in distributions that are not significantly multi-modal. We validate this method through visual inspection, verifying that PDFs with zero inflection points are actually very flat and uninformative, and those with more than one are significantly multi-modal. Therefore, we keep those PDFs with only one convex inflection point. Throughout the paper, we refer to this specific flag as \texttt{unimodal}.

    To determine whether a posterior is very skewed toward the parameter space boundary, the most intuitive method is to examine whether the estimated upper and lower bounds exceed the parameter space. 
    Here, we adopt a more flexible approach based on the skewness computed as the adjusted Fisher-Pearson standardized moment coefficient of the PDF: if the absolute value of the skewness is $>1$, the chain is considered highly skewed or asymmetric, indicating that it is converging toward the edge of the parameter space. We validate this algorithm by visually inspecting the one-dimensional posterior PDFs of a fully fitted catalogue. Throughout the paper, we refer to this specific flag as \texttt{not truncated}. 

    In our Bayesian analysis, there are numerous free parameters involved, including the model components of noise and dust as discussed in Section~\ref{sec:method} for which in principle one may apply these automatic inspection techniques. In this work, we consider only four key parameters associated with galaxies' physical properties, namely, age, $\tau$, $Z$, and $M_\star$. The final \texttt{unimodal} (or \texttt{not truncated}) flag is then taken from the intersection (AND logic) of the individual flags obtained from all four parameters.

    We find that the \texttt{not truncated} flag removes more galaxies in the lower age regime, by increasing the median of the age from $\left\langle t\right\rangle=4.1\pm1.5$ to $5.6\pm2.4\ \mathrm{Gyr}$ and the value of $\tau$ from $\left\langle \tau \right\rangle=0.5\pm0.3$ to $0.7\pm0.3\ \mathrm{Gyr}$ respectively, while the \texttt{unimodal} selection negligibly affects the shape of the posterior distribution of the derived parameters, as well as their median values. Though potentially improving its reliability, the \texttt{not truncated} and \texttt{unimodal} flags shrink the sample by approximately 20\% and 50\%--60\%.

\begin{figure}[t]
    \centering
        \includegraphics[trim={0.4cm 0 0 0}, width=1.1\hsize]{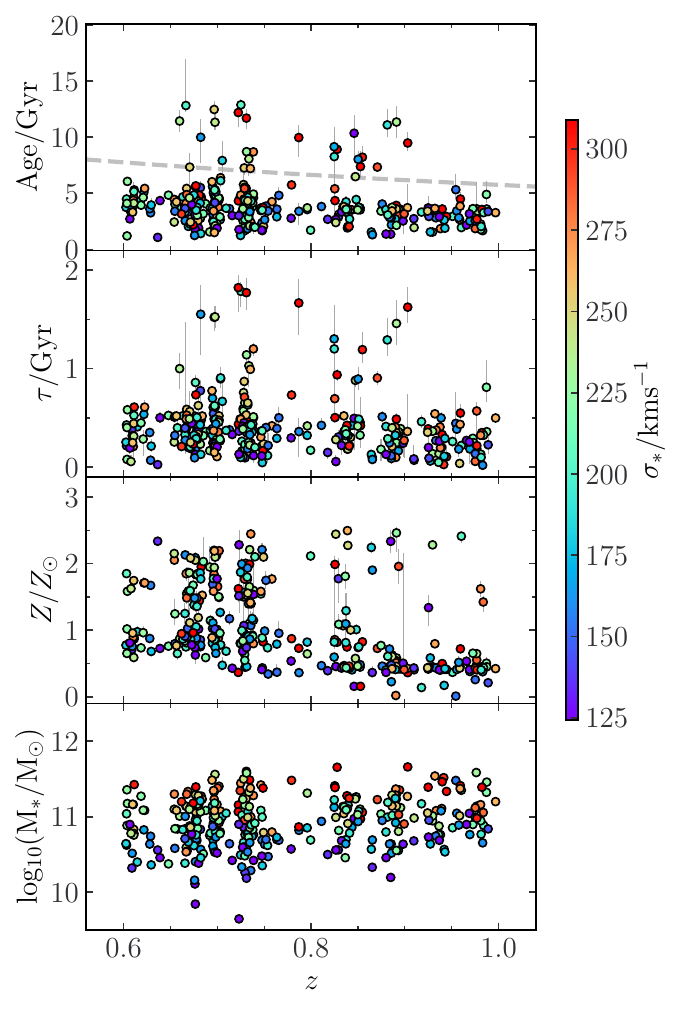}
        \caption{Distributions of stellar ages, star formation timescales ($\tau$), metallicities ($Z$) and logarithmic stellar masses ($\log(M_\star/M_\odot$)) versus $z$ obtained from the full spectrum fitting of 335 individual cosmic chronometers in LEGA-C DR2. Each galaxy is color-coded by its stellar velocity dispersion ($\sigma_\star$). The first two and the last panel span the entire parameter space explored (for a more detailed description, see Section~\ref{sec:method}.
        \label{fig:individuals}}
    \end{figure}

    \paragraph{\bf Poorly modelled CaII~H/K} 
    The CaII~K and H lines, centered respectively at $3934$ and $3969$~{\AA} rest frame (see Figure~\ref{fig:individuals}), are two prominent features in galaxy spectra. In galaxies dominated by an old stellar population, it is usually found that the K line is deeper than the H line, this being opposite in presence of young star-forming components (see figure~5 in \citealt{Moresco2018ApJ}). In our analysis, we adopt the definition of H/K introduced by \cite{Fanfani2019} that consists in measuring the ratio of two pseudo-Lick indices CaII~K and H, i.e. $\mathrm{H/K}=I_\mathrm{H}/I_\mathrm{K}$. This technique is less sensitive to potential bias introduced by noise peaks in the spectrum with respect to using the H and K flux minima, i.e. $|\mathrm{H/K}|_{min}=F_{min}(\mathrm{H})/F_{min}(\mathrm{K})$. 

    This diagnostic has been used to test the presence or absence of a contaminant population, therefore describing the purity of the selected sample \citep{Moresco2018ApJ, Borghi2022a}. In particular, \citet{Borghi2022a} found that $\mathrm{H/K}<1.2$ is safely equivalent to $|\mathrm{H/K}|_{min}>1$, and $\mathrm{H/K}<1.1$ well reproduces other selection criteria including NUVrJ \citep{Ilbert2013} and $\mathrm{sSFR/yr}<-11$ cut. They tested that the current sample of passive galaxies has a typical value of $\mathrm{H/K}=0.96\pm0.08$, validating the purity of the selection. We, as well,  do not observe any correlation between galaxies' properties (especially age) and the $\mathrm{H/K}$ in both the observed and posterior spectra, indicating that no significant contribution from young stellar component is present in our passive galaxies sample. This also excludes the possible presence of galaxies that experienced recent rejuvenation events as found in \cite{Chauke2019ApJ}.

    It is, therefore, plausible to consider that the posterior spectra which do not adequately reproduce this feature are not completely appropriate fits, and should be excluded. We calculate the $\mathrm{H/K}$ ratios in the observational LEGA-C DR2 spectra and in our posterior spectra using \pylick\footnote{The code is available at \url{https://gitlab.com/mmoresco/pylick/}}. Then, we associate a flag with removing galaxies from the sample according to the following criterion: if the discrepancy between the observed and the inferred value is greater than 10\%,
    \begin{equation}
        \frac{|\mathrm{H/K}_\mathit{post}-\mathrm{H/K}_\mathit{obs}|}{\mathrm{H/K}_\mathit{obs}}>10\%,
    \end{equation}
    we consider that the feature has not been well reproduced, resulting in a potential deviation in the age estimation of the galaxy. We define the flag corresponding to this criterion \texttt{H/K well modelled}. Applying this flag, we observe a considerable reduction in the long-tail shapes of the age and $\tau$ distributions in Figure~\ref{fig:hist_posteriors}, which clearly demonstrates the importance and validity of this criterion. 

    \paragraph{\bf Inconsistency between Photometric and Spectroscopic Data} 
    The calibration of spectroscopic and photometric data is a complex process, and systematic differences between data obtained using different calibration pipelines are possible. To ensure a proper combination of the different sets of data, we must compare the photometric points (when available) in the wavelength region covered by the observed spectrum to the flux of the spectrum at the effective position of the corresponding photometric filters(see description in Section~\ref{sec:data});
    if the difference between the two is significant, it means that there is an inconsistency between the photometric and spectroscopic points, and therefore these data cannot be fit jointly.
    We define the median absolute pull (MAP) between the photometric fluxes and the spectral fluxes in the form of 
    \begin{equation}
        \mathrm{MAP} = \mathrm{Median}\Bigg(\frac{{\bigg|{f_{spec,i}}-{f_{phot,i}}\bigg|}}{\sqrt{{\sigma}^2_{f_{spec,i}}+\sigma^2_{f_{phot,i}}}}\Bigg),
    \end{equation}
    where the spectral fluxes are averaged over extremely narrow wavelength windows (10 pixels $\sim 6$~{\AA}) centered at the effective wavelength of the filters. We conservatively keep only the galaxies with a $\mathrm{MAP}\leq1$. Such flagging is stringent enough to rule out most of the largest potential deviations. Throughout the paper, we refer to this specific flag as \texttt{better SEDs}. Applying this flag, we observe the median of the age and $\sigma_{\star}$ are equivalent to the values when applying the \texttt{H/K well modelled} flag, even though not significant, suggesting the inconsistency has a negative effect on simultaneously fitting spectroscopic and photometric data. 

    Inspecting Figure~\ref{fig:hist_posteriors}, we notice how in the case when only the spectrum is fitted, the various flags defined are crucial to significantly help reduce the tails of nonphysically large ages ($age>7$ Gyr), and as well the largest values of $\tau$ ($\tau>1$ Gyr), that were galaxies severely affected by the age-$\tau$ degeneracy fitted with a larger value of age and $\tau$. This effect is in particular evident by comparing the median values of each distribution as a function of the different flags applied, as also reported in the figure. At the same time, analyzing the distribution of the fit obtained from the joint analysis of spectrum and photometry, we notice how the impact of the flagging is significantly smaller, and that the addition of the photometric bands per-se helps in reducing the degeneracy between parameters and obtaining well-converged fits, with an extremely negligible fraction of points at high ages and $\tau$. It remains also interesting to notice that despite wide and uninformative priors on age and $\tau$, the derived best fits confirm the fact that these objects have been selected extremely accurately, and that they are old objects formed over relatively small time-scales.

    In conclusion, we decide to keep as our baseline for the analysis the fit that includes spectroscopic and photometric data without including any flagging, maximizing in this way both the accuracy of the results and the final statistics. In Section~\ref{sub:measuring_hz} we will discuss and quantify the impact of applying these flags on our cosmological result.

    \subsection{Galaxy Properties} \label{sub:galaxy_properties} 

    Figure~\ref{fig:individuals} presents the final constraints to the stellar population properties, namely the best-fit stellar age, metallicity ($Z$), mass ($M_{*}$), and star formation timescale ($\tau$), by assuming a delayed exponentially declining star formation history (see the last row of Table~\ref{Table:Priors} for more details on the model ingredients and adopted priors). We use the light-weighted properties instead of the mass-weighted ones, considering that the latter are more sensitive to the choice of the star-forming history parametrization \citep{Conroy2013ARAA}. Each galaxy is color-coded by its stellar velocity dispersion. Interestingly, even if do not bind the upper age value with a cosmological prior, we find that this population of galaxies qualitatively follows the descending trend predicted from the $\Lambda$CDM model assuming the \cite{Planck2020AA} parameters, with less than 10\% of galaxies higher than the reference $\Lambda$CDM boundary. The median age is $\left\langle t\right\rangle=3.60\pm0.82$~Gyr. For comparison purposes only, when assuming a baseline $\Lambda$CDM model, this value corresponds to a typical formation time of $\sim3$~Gyr after the Big Bang, or $z_f\sim 2.5$, in agreement with a wealth of literature data \citep[e.g.][]{Gallazzi2014, Carnall2019MNRAS, Belli2019, Carnall2022, Tacchella2022}. In Section~\ref{sec:hubble_parameter} we further study the $age(z)$ relation and its trends with $\sigma_\star$. 

    For the first time, we are able to quantitatively study the star-formation timescale $\tau$ -- redshift relation for this sample of cosmic chronometers. Even if we adopt a wide Uniform prior on $\tau\sim\mathcal{U}[0, 2]$~Gyr, we find typical $\left\langle \tau\right\rangle= 0.36\pm0.13$~Gyr, with about 80\% of the galaxies having $\tau < 0.5$~Gyr. Most importantly, from this analysis we find no significant dependence on $z$. This is an additional confirmation that the current sample of passive galaxies is very homogeneous in its physical properties over different $z$.  In addition, we do not find a statistically significant trend of $\tau$ with $\sigma_\star$. This is not in contradiction with the idea that more massive galaxies formed in shorter timescales (as expected from the downsizing-scenario; \citealt{Cowie1996}). On the contrary, we are selecting the very massive and passive envelope of objects, so that we expect shortest $\tau$ and no correlation with mass. This may also explain the reason why our SFHs are shorter the those derived by \cite{Chauke2018} on LEGA-C quiescent galaxies. But for a definitive answer, the impact of different SFH assumptions must be further assessed. Similar values of $\tau$ for quiescent galaxies at intermediate redshifts were also obtained by \cite{Pacifici2016}, \cite{Carnall2019MNRAS}, and \cite{Tacchella2022}

    We find stellar metallicities with slightly undersolar values, $\left\langle Z/Z_\odot\right\rangle= 0.84\pm0.41$, in agreement with \citet{Borghi2022a} who find typical $Z/Z_\odot\sim 1.1$ using Lick indices. This result further strengthens the idea that it exists a population of massive and passive galaxies which, at least up to $z\sim 0.8$, does not evolve significantly in its metal content and has values similar to those of their local counterparts \citep[see also ][]{Thomas2011, Gallazzi2014, Onodera2015, EstradaCarpenter2019}. However, we observe an evolution toward smaller $Z$ with increasing redshift as observed in \cite{Beverage2021} for 68 massive quiescent LEGA-C galaxies, or in \cite{Carnall2022} with VANDELS galaxies at $z\sim 1.2$. We traced that this effect is due to a degeneracy between metallicity and dust in the fit, because it completely disappears when we remove that parameter from the fit. We notice also that, instead, the differential age measured is very stable, since the Hubble parameter derived in that configuration varies only by 1.1\% with respect to our baseline, well below the currently estimated error. We discuss this point in Appendix A.

    The stellar masses derived in the analysis correlate with the observed stellar velocity dispersion. This is a well-established result, usually interpreted with the idea that galaxies with a larger gravitational potential well are capable of retaining more gas and therefore forming more stars. 

    In conclusion, our sample of cosmic chronometers shows ages and star-formation timescales supporting the scenario that they must have formed at early epochs and with very short star formation events quickly exhausting their gas reservoir and then evolved passively.

\section{From Differential Ages to the Hubble Parameter} \label{sec:hubble_parameter}

\subsection{Binning Parameters} \label{sub:binning_parameters}
    To apply the CC approach, we need to derive from the $age(z)$ relation obtained in our analysis the differential age evolution $\Delta t$ in a given redshift bin $\Delta z$. Since this measurement involves the estimation of a derivative, it is typically convenient, in case of noisy data, to increase the S/N of the data by averaging different values, having a consequently more robust estimate of the differential age. This same approach has been adopted in most of the CC studies, see e.g. \cite{Moresco2012, Moresco:2015, Moresco:2016, Borghi2022b}.

    Following the previous works by \cite{Moresco:2016, Borghi2022b}, we decide to average our data not only as a function of redshift, but also as a function of the velocity dispersion; this last step is in particular important on the one side since it allows us to detect possible trends of the physical parameters as a function of $\sigma_\star$ (i.e. as a function of the stellar mass), but at the same time because, as highlighted by \cite{Thomas2011} stellar populations of different stellar mass correspond to population formed at different times and over different timescales. Performing an analysis at almost constant velocity dispersion (or stellar mass) ensures the homogeneity of the tracers compared and as a consequence a non-biased determination of the Hubble parameter. For more details, see \cite{Moresco2022}.

    We note here that before binning and averaging our data, we further excluded 15 objects from our sample since they had a redshift significantly different from the bulk of the population (see Figure~\ref{fig:hist_posteriors}); we, therefore, imposed a cut $0.6\leq z\leq 1.0$, ending up with 335 galaxies. From now on, all the results will be referred to this sample.

    \begin{figure}[htb!]
        \includegraphics[width=.95\hsize]{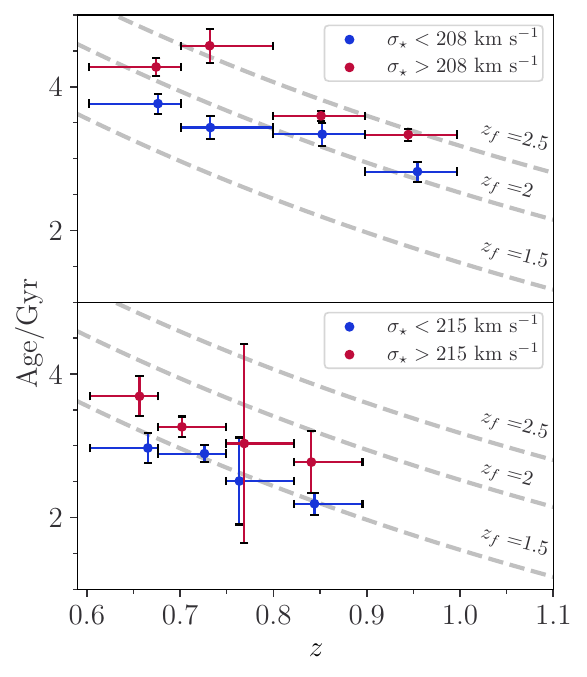}
        \vspace{-1em}
        \caption{Median binned age-redshift relations for our baseline results based on full-spectrum fitting (top) and for the Lick indices analysis from \cite{Borghi2022b} (bottom). The blue and red points represent the lower and higher $\sigma_\star$ bins, divided using the median value of each sample as the threshold. For each bin, the vertical error bars are the errors associated to the median ages, 
        while horizontal bars denote the bin width. For illustrative purposes only, we include the redshift of formation (gray dashed lines) assuming a reference $\Lambda$CDM model from \cite{Planck2020AA}.}\label{fig:general}
    \end{figure}

    Several different choices of binning can be adopted, including the type and number of bins, as well as the type of average statistics adopted. In particular, we can choose to divide the data into $N\times M$ bins of redshift and velocity dispersion, to have bins of fixed widths or divided into equi-populated quantiles, and to estimate, within each bin, the averaged quantities with different methods (mean, median, weighted mean). We must consider a trade-off between the benefit of avoiding uneven data distribution by using quantile bins and the benefit of improved population separation by using fixed bins. Additionally, we must use a sufficient number of bins to achieve both good statistics and an optimal sampling of the age-$z$ trend. If the number is too large, the statistics of each bin deteriorate and create large oscillations; on the other hand, if the number is too small, the evolution trend is smoothed out.

    We consider median statistics instead of arithmetic (or weighted) average because it is less sensitive to outliers (including badly constrained galaxies, see Section~\ref{sub:breaking_degeneracies}. In particular, we decided to avoid the use of weighted average  because we observed a positive correlation between the estimated ages and their uncertainties, with younger galaxies having smaller uncertainties. Therefore, by using a weighted average, we would have biased our result toward smaller ages. We use the sampling error of the median in the form of $\sigma=\mathrm{MAD}/\sqrt{N}$, where $\mathrm{MAD}$ is the median absolute deviation and $N$ is the number of galaxies in that bin. 

    We test all the possible choices for binning, including combinations of quantile and fixed binning for $z$ and $\sigma_\star$, combinations by taking reasonable values from \{4, 2, 1\} and $z$ in \{2, 1\} for $N_{z \mathrm{bin}}$ and $N_{\sigma_\star \mathrm{bin}}$ respectively. We do not use additional bins to ensure that each bin contains a sufficient number of galaxies. We have approximately an average of 40 galaxies per bin in the case of 4$\times$2 bins that significantly decreases to 20 when the $H/K$ selection is applied. After a careful comparison of the various options, we decide to consider as our baseline binning $4$ fixed $z$ bins combined with $2$ quantile $\sigma_\star$ bins and median statistics, since it provides the best trade-off between a large enough redshift range probed, a separation in velocity dispersion allowing us to study the mass effect, and the same number of bins to compare our results to \cite{Borghi2022b}.
    We discuss the rationale of this choice in further detail, as well as quantify its impact on the results, in Section~\ref{sub:measuring_hz}. 

    In Figure~\ref{fig:general}, we compare our baseline binned $age(z)$ to the one obtained by \cite{Borghi2022b}. As a first point, we observe an offset between the absolute ages estimated in the two methods of about $0.61\pm0.05$ Gyr. This difference can be actually explained and interpreted by taking into account the different SFH adopted in the two analyses. In \cite{Borghi2022b}, the theoretical models for the Lick indices where available only for SSP, while in this analysis, we assumed a more complex and realistic SFH, which is also one of the improvements of this analysis with respect to the previous one. The net effect is the bias in age observed. 

    It is however striking to observe the accuracy with which both derived ages evolve as a function of redshift. Qualitatively, they both follow extremely well the cosmological lines reported as a reference in the figure, demonstrating that despite the difference in the method, the difference in the assumed SFH, the slightly different threshold in $\sigma_\star$ adopted, the different number of objects, the differential ages agrees extremely well. This will be demonstrated quantitatively in the following section, but it is important to stress here that this is the first time that two different methods to derive differential ages on a common sample of pure massive and passive galaxies (cosmic chronometers) have been performed. The agreement found between the trends of $\Delta t$ is, therefore, an additional piece of evidence supporting the robustness of the CC method as a cosmological probe.

    Additionally, we observe a distinct mass-downsizing pattern for which more massive galaxies ($\sigma_\star >208\ \mathrm{km s^{-1}}$) exhibit a higher redshift of formation ($z_f\sim2.5$) with respect to less massive ones ($z_f\sim2$). Our $z_f$ estimates are approximately $0.5$ higher than those in \cite{Borghi2022b}, still due to the use of a more extensive and realistic SFH (DED model), that allows the star formation to start earlier and persist over a longer period of time with respect to the SSP. Our results also achieve higher $z$ and smaller errors of binned ages due to the use of a larger sample of galaxies (335) with respect to \cite{Borghi2022b} (140), which reduces statistical errors, particularly in the third $z$ bin. With the same number of $z$ bins, our result also gives a larger range of redshift, allowing us to probe to Universe up to a slightly higher $z$.

    \subsection{Measuring \ensuremath{H(z)}}\label{sub:measuring_hz}

    We compute $H(z)$ by using binned ages and redshifts described in the previous section,
    \begin{equation}
    \label{eq:Principle}
        H(z) = -\frac{1}{(1+z)}\frac{\Delta z}{\Delta t}.
    \end{equation}
    To minimize the impact of fluctuations in the data, we do not use consecutive bins to calculate $\Delta z$ and $\Delta t$, but a difference approach based on non-adjacent bins, estimating the difference, in a given $\sigma_\star$ bin, between the i$^\mathrm{th}$ and the (i+N/2)$^\mathrm{th}$ point, where N is the number of redshift bins defined. As presented in Section~\ref{sub:binning_parameters} we use two types of bin type: fixed and quantiles (i.e. flexible to ensure an equal number of objects in each bin). This strategy requires making an even number of redshift bins to avoid covariance between results caused by the multiple uses of the same data point. This approach allows us also to estimate the difference between points where the expected age evolution is larger than the associated error, making the estimate of $\Delta t$ less noisy and more robust. This differential approach of the cosmic chronometer method plays a crucial role in minimizing the rejuvenation in the star-formation history \citep{Moresco2022}.

    We perform this evaluation in each $\sigma_\star$ regime, obtaining $\left( N_{\mathrm{bin}, \sigma_\star} \times N_{\mathrm{bin}, z}\right) /2$ Hubble parameter measurements. Finally, these values are combined to get a single and more accurate estimate of $H(z)$ with an inverse-variance weighted average \citep[as also done, e.g. in][]{Moresco:2016, Borghi2022b}.

    The different choices of how to bin and select our data, as well as the assumed SFH, are potential sources of systematical uncertainties for the final $H(z)$. In our analysis, we do not account for other systematics introduced by other assumptions of Stellar Population Synthesis (SPS) model, as we discuss in Section~\ref{sec:conclusions} \citep[for a detailed treatment, see][]{Moresco2020ApJ, Moresco2022}. In summary, starting from our baseline result, we estimate the impact on the cosmological results by adopting: different choices of binning (Section ~\ref{sub:binning_parameters}), different flagging methods (Section~\ref{sub:breaking_degeneracies}), or a different SFH assumption that decouples the rising and declining slopes of the SFH (i.e., double power-law or DPL, see Section~\ref{sec:method}).  This will allow us to estimate the systematic errors due to these effects to be associated to our measurement. The results are shown in Figure~\ref{fig:systematics}. 

    First of all, we observe that the configurations contributing the most to a systematic difference in $H(z)$ are those in which all galaxies  are averaged together in a redshift bin independently of their $\sigma_\star$. The larger shift underlines, even more, the need to perform the analysis of CC carefully selecting the sample in bins of velocity dispersion (or stellar mass), otherwise, the assumption of having a homogeneous sample of chronometers is dropped, and mixing different galaxy populations will exacerbate the progenitor bias \citep{Dokkum2000ApJ}. As shown in Figure~\ref{fig:systematics}, we end up smoothing the evolutionary trend, obtaining a higher $H(z)$ and a larger scatter, resulting in larger statistical uncertainties. We also note that by using (equi-populated) quantile bins in $z$ we obtain higher $H(z)$ values with respect to fixed bins. This behavior may be explained by the uneven redshift distribution of our galaxies (see Figure~\ref{fig:hist_sample}). In particular, because there are fewer high-redshift galaxies, the high-redshift quantile bins span a much wider interval, thus flattening the $age(z)$ relation and, ultimately, increasing $H(z)$ and its associated uncertainty. On the contrary, the $\sigma_\star$ distribution is approximately Gaussian, which makes the result of quantile and fixed $\sigma_\star$ bins not significantly different. 
    
    As for the number of $z$ bins, the results show that a smaller or larger number of bins produce a higher $H(z)$, which is reasonable because,  due to the redshift distribution, a smaller number of bins does not allow to correctly map the slope of the age redshift relation, since the larger $dz$ and lower statistics at high $z$ would artificially flatten the median $age(z)$. Besides, taking more bins, the data will be noise dominated.

    Even-though flagging posteriors slightly changes the value of $H(z)$, all the related results are compatible with our baseline result as shown in Figure~\ref{fig:systematics}.

\begin{figure}[ht]
        \centering
        \includegraphics[trim={1 0.5cm 0 0 0}, width=\hsize]{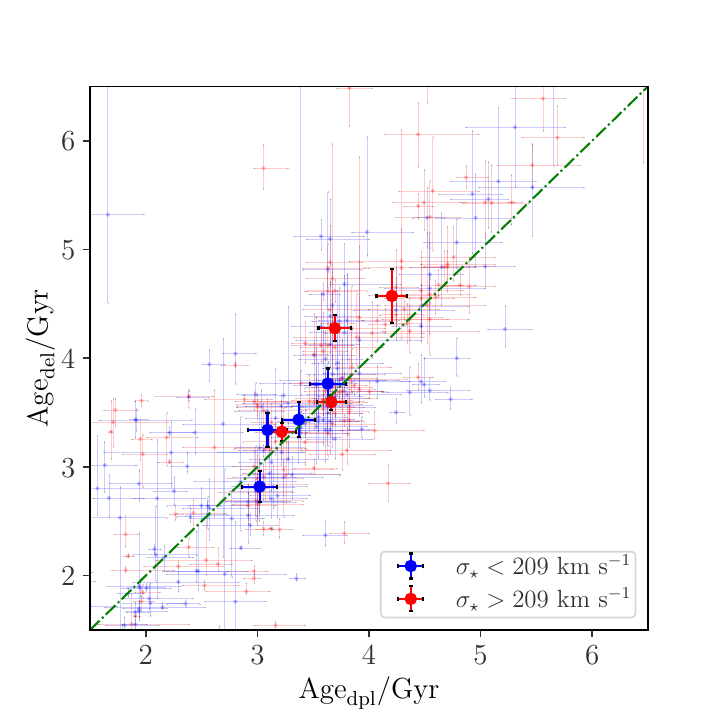}
        \vspace*{4.8em}
        \caption{Galaxies' ages fitted by a full spectroscopic and photometric fitting based on the delayed exponentially declining model versus those fitted under the same fitting configurations but based on the double power law model. The thin error bars represent measurements of individual massive and passive galaxies selected in LEGA-C DR2, while the thick error bars are the binned results (obtained with the 4 fixed $z$ bins $\times$ 2 quantiles $\sigma_\star$ bins), which are both colored by their $\sigma_\star$ with lower bins blue and higher bins red.
        The green dashed-dotted line marks out the diagonal direction.}
         \label{fig:age_dpl_ded}
    \end{figure}

\begin{figure*}[t!]
        \includegraphics[width=1\textwidth]{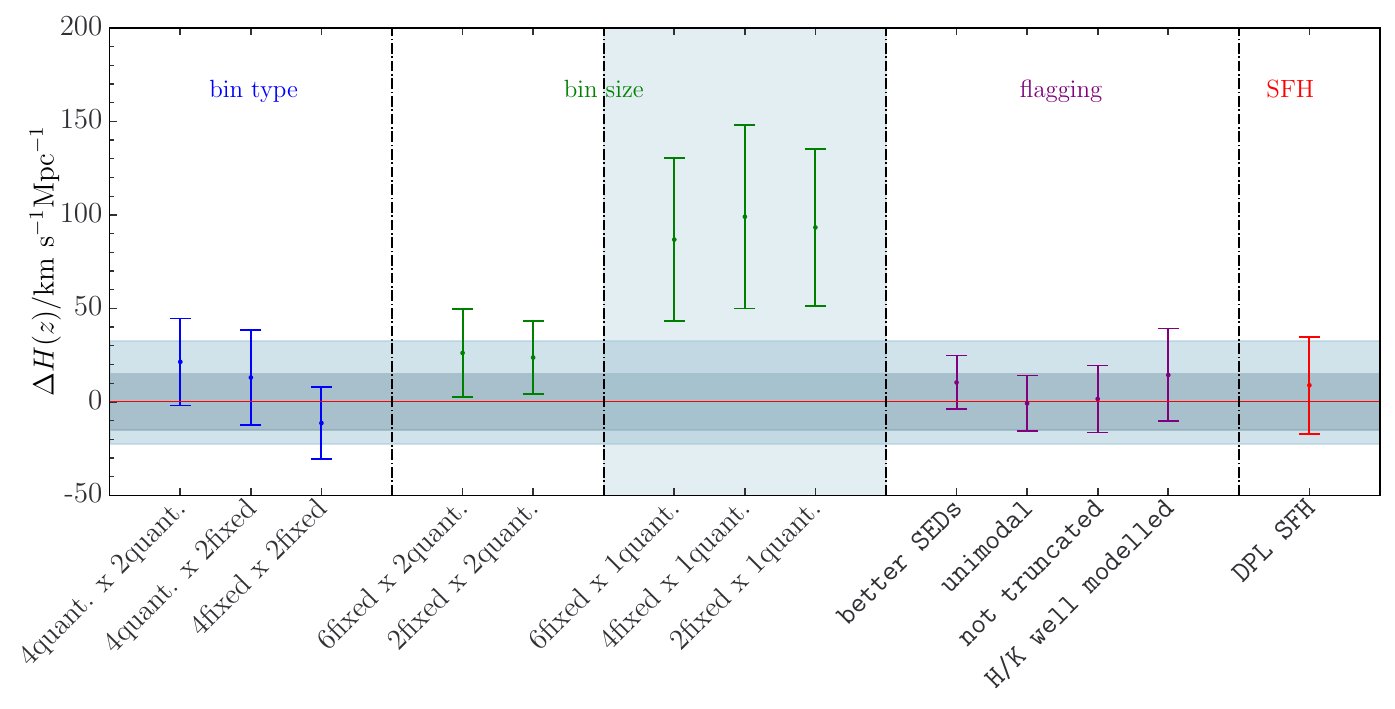}
        \caption{Difference between our final $H(z)$ measurement (obtained with 4 fixed $z$ bins $\times$ 2 quantiles $\sigma_\star$ bins) and the other values obtained by varying the bin settings, in particular the bin type (blue), and the bin size (green),  by applying different quality flags (violet) as described in Section~\ref{sub:breaking_degeneracies}, or by fitting an alternative double power-law SFH (red). The horizontal dark and light-shaded region represent the statistical and total uncertainty of the final $H(z)$, the latter defined as $\sqrt{\sigma_{\mathrm{stat.}}^2+\sigma_{\mathrm{syst.}}^2}$. The values obtained without binning in $\sigma_\star$ (vertical shaded region) are the most significant contributors of the systematics error budget.} \label{fig:systematics}
    \end{figure*}

    The various choices of the SFH could further contribute to systematic uncertainties in the measurement of  $H(z)$. To address this point, we fit our sample with a different more flexible SFH model, DPL  (see section~\ref{sec:method}), commonly used in other BAGPIPES analyses \citep{Carnall2018MNRAS}.  To evaluate the difference brought on by a change in the SFH, we solely alter the SFH assumed and keep the other fitting configuration unchanged. In Figure~\ref{fig:age_dpl_ded}, we make a direct comparison between the ages obtained using the two models, showing that the ages estimated with the two models are compatible. To quantitatively assess the impact of choosing a different SFH on our result, we estimate the Hubble parameter with the DPL SFH. With all other configurations of analyses unchanged, we obtain a $H(z=0.80) = 122.0 \pm 21.1~\mathrm{km~s^{-1}~Mpc^{-1}}$ using the results by fitting the DPL model, showing a 7.9\% difference comparing to our baseline $H(z)$. We take this as an estimation of the systematic uncertainty caused by choice of the SFH model.
    
    To assess the systematic error, we, therefore, consider our baseline result and quantify how much the results are perturbed by three sources of systematics, namely varying the binning scheme, the applied quality flags, and by assuming a different double power law SFH. We estimate the median difference between our baseline $H(z)$ and the measurements obtained from each source, taking
    \begin{equation}
        \sigma_{\mathrm{syst.} j} =  \mathrm{median}\Big(H(z)_j-H(z)_{\mathrm{base}}-\Delta H_{\mathrm{model}}\Big).
    \end{equation}
    In this equation, we need to account for the difference caused by redshift evolution, $\Delta H_{\mathrm{model}} = H_{\mathrm{model}}(z_{\mathrm{base}}) - H_{\mathrm{model}}(z_j)$, where the choice of the assumed cosmological model negligibly affects in the case of minuscule redshift difference. We compute the total systematic uncertainty by summing each contribution in quadrature and calculating the upper and lower $\sigma_\mathrm{syst. tot.}$ separately.

    \begin{figure}[t]
        \includegraphics[trim={1 0.5cm 0 0 0}, width=\hsize]{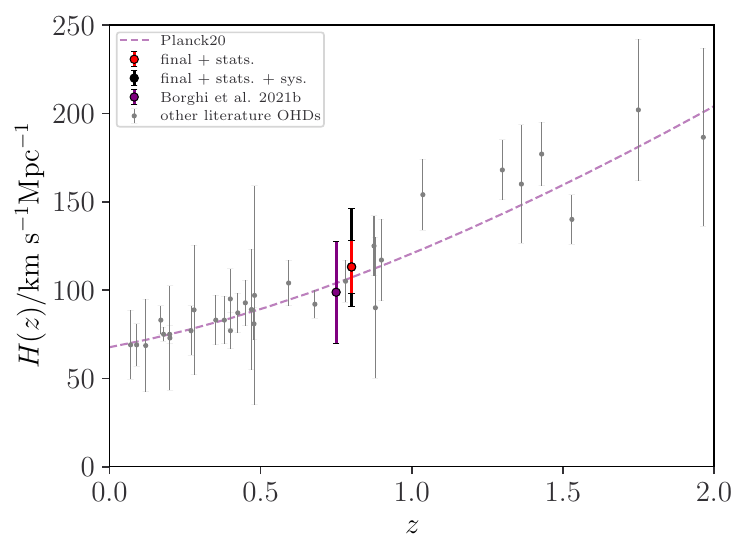}
        \caption{Final $H(z)$ value obtained from full spectral fitting of 335 cosmic chronometers in LEGA-C DR2 with statistical (red) and total (black) uncertainty. The violet point is the value obtained by \cite{Borghi2022b} via Lick indices analysis of 140 galaxies of this dataset. The gray points are all the other OHDs available from the literature \citep{Jimenez:2003,Simon:2005,Stern:2010,Moresco:2012,Zhang:2014,Moresco:2015,Moresco:2016,Ratsimbazafy:2017}. For illustrative purposes only, we include the $H(z)$ prediction assuming a $\Lambda$CDM model with \cite{Planck2020AA} parameters.}\label{fig:final}
    \end{figure}

    In summary, we obtain a new cosmology-independent measurement of $H(z)=113.1 \pm15.1(\mathrm{stat.})^{+29.1}_{-11.3}(\mathrm{syst.})$ $\mathrm{km\ s^{-1}\ Mpc^{-1}}$ at $z=0.80$. Our measurement is consistent with the $H(z=0.75) = 98.8 \pm 33.6 \ \mathrm{ km\ s^{-1}\ Mpc^{-1}}$ that \cite{Borghi2022b} obtained. We notice that the statistical error decreases from $24.8 \mathrm{ km\ s^{-1}\ Mpc^{-1}}$ to $15.1 \ \mathrm{ km\ s^{-1}\ Mpc^{-1}}$, i.e. by a factor of approximately 0.6, equivalent to the inverse of $\sqrt{335/140}$ due to the increasing number of cosmic chronometers used for our $H(z)$ measurement. The biggest contribution to this uncertainty is given by the binning scheme, and in particular when H(z) is computed without separating the galaxies into two $\sigma_\star$ subsamples. By excluding this contribution the upper systematical error decrease to $26.1~\mathrm{ km~s^{-1}~Mpc^{-1}}$. In Figure~\ref{fig:final}, we compare our final result with all the currently available $H(z)$, finding a noticeable consistency.

\section{Conclusions}\label{sec:conclusions}

    In this paper, we analyze a sample of 350 massive and passive galaxies mostly (95\%) at $0.6\lesssim z\lesssim1.0$ extracted from LEGA-C DR2 in \cite{Borghi2022a}, deriving their physical properties from a full-spectral fitting analysis. Given that the ultimate goal of this work is apply the CC method with this dataset, it was optimized and thoroughly tested to minimize the possible contamination of the sample by young, star-forming objects. The derived $age(z)$ relation is then used to constrain the differential ages $\mathrm{d}t$, and to provide a new estimate of the Hubble parameter $H(z)$. Our analysis will also allow us to compare for the first time on the same dataset the differential ages $\mathrm{d}t$ derived with two different and independent methods, namely Lick indices and full-spectral fitting, and to investigate and validate the capability of the CC approach to robustly derive $H(z)$.

    Here, we utilize the public code \bagpipes~ \citep{Carnall2018MNRAS} to derive stellar ages, metallicities, and SFH fitting both the spectroscopic data alone and the spectroscopic and photometric data jointly for all the individual galaxies in our sample. Our main results are summarized as follows.

    \begin{itemize}
        \item We first extend \bagpipes~ by removing the cosmological prior on the derived ages, to avoid possible biases on our cosmological results due to the assumption of a fiducial cosmology. We also adopt flat uninformative priors on all the derived quantities, namely the stellar age, metallicity, and SFH. We then explore the dependence of our results on the SFH assumed, the parameters included in the fit, and the priors considered. We opt to use a delayed exponentially declined SFH, improving with respect to the analysis of \cite{Borghi2022a} where SSP where assumed, but minimizing the number of free parameters for the functional form of the SFH, since we verified that the gain in the quality of the fit was marginal with other choices.
        \item We find the results obtained from the fit to the spectroscopy alone to be less accurate and more scattered than the results obtained by fitting the combination of spectroscopy and photometry, with in particular larger tails toward higher ages and $\tau$. We defined a set of indicators of quality of the fit and convergence criteria, based on the inspection of the posterior distribution and on the accuracy with which the best-fit was reproducing specific observational features in the spectrum, namely the CaII H/K known to be correlated with potential episodes of recent star formation. We demonstrate that the nonphysical scatter in the derived parameter obtained using only spectroscopic data can be lifted by applying masks defined on these indicators. 
        \item We observe that the inclusion of photometric data (21 bands in this analysis) allows the fit to converge correctly even without applying the convergence criteria previously discussed, reducing the degeneracy between parameters. As a consequence, in this framework we are able to maximize the final number of objects with a correct fit, and we decide to consider this as the baseline of our analysis.
        \item We find that the measured $age(z)$ relation are well compatible with a cosmological aging as a function of redshift even with assumed a flat prior on age$\in\mathcal{U}(0,20)$ Gyr. Our results present also a clear downsizing trend when divided into two bins of velocity dispersion, with galaxies with $\sigma_\star>108$ km s$^{-1}$ with a formation redshift $z_{f}\sim2.5$ and the ones with $\sigma_\star<108$ km s$^{-1}$ with a formation redshift $z_{f}\sim2$. Even though we consider a significantly wider prior in our analysis, these galaxies show very short star formation timescales with a median value of $\left\langle \tau\right\rangle= 0.36\pm0.13$, 
        \item The average measured stellar metallicity is $\left\langle Z/Z_\odot\right\rangle= 0.84\pm0.41$ with a small hint of evolution as a function of redshift. We prove that this tension or fitted evolution is due to the degeneracy between the metallicity and dust in the fit, and demonstrate that it has an almost negligible effect on the differential age.
        \item We compare for the first time the stellar ages derived from two very different and independent methods applied on the same sample, the full-spectrum fitting and the Lick indices analysis. We find that the absolute ages derived present an offset of $0.61\pm0.05$ Gyr, which can be understood by recalling that the Lick index models assume an idealized SSP SFH that slightly bias the absolute values toward younger ages. The agreement, however, between the differential ages is striking, and it is important to underline here that the CC approach is based on the measurement of $\mathrm{d}z/\mathrm{d}t$, that perfectly agrees within errors in the two measurements. This result is in particular important because it demonstrates the robustness of the method and the stability of the $\mathrm{d}t$ measurement, confirming that it can be derived with significantly less biases than absolute ages.
        \item From the analysis of the binned $age(z)$ relation, we derive a new $H(z)$ measurement $H(z=0.80)=113.1\pm15.1(\mathrm{stat.})$. We verify that our result is fully compatible with the one by \cite{Borghi2022b}, even if at a slightly larger redshift and with a slightly smaller statistical error due to the different number of objects in the final sample used by the two methods \citep[in][to ensure homogeneity in the analysis we decided to consider in the final sample only the spectra for which the same number of spectral features where observable]{Borghi2022b}. we also test that our result is consistent with other literature OHDs, and as well with the prediction of $\Lambda$CDM model assuming the \cite{Planck2020AA} cosmological parameters.
        \item We assess the systematics involved in the results by varying the methods with which the binned $age(z)$ relations are derived, changing the number of bins, the method to estimate the average value, the assumed SFH model in the fit, and testing the application of the masks described above. We estimate a systematic error of ${+29.1}\;{-11.3}(\mathrm{syst.})\ \mathrm{ km\ s^{-1}\ Mpc^{-1}}$, mainly dominated by the large variation in the results obtained when the sample is not divided into two $\sigma_\star$ (or stellar mass) bins, suggesting that an analysis in specific ranges of masses is fundamental to ensure the homogeneity in formation of the CCs considered.
        \item In the end, we obtain a measurement of the Hubble parameter $H(z)=113.1 \pm15.1(\mathrm{stat.})^{+29.1}_{-11.3}(\mathrm{syst.})$ at $z=0.80$.
    \end{itemize}

    We underline that since the sample used in this analysis and in the one of \cite{Borghi2022b} are drawn from the same parent sample, it should be avoided to use them in combination, since the measurements will be highly covariant. We also underline that in the current analysis we decided not to explore the further dependence of our result on other assumptions of the SPS models. Regarding the dependence on different SFH, we verified that within \bagpipes~ the SFH choice is somehow limited (we could have chosen amongst SSP, DED, and DPL, but the SSP yields discrete pattern on galaxies' ages, weakening the reliability of the  differential ages obtained, which however can be improved by a re-run of the SPS on more refined grids.), therefore it would not have allowed a full estimate of this effect. Moreover, the fit obtained with the DPL SFH provided similar results to the ones obtained here, but with a higher number of free parameters, not justifying in our case the choice of that SFH. In the end, we chose the DED model as our baseline, and the DPL for evaluating the SFH choice caused systematic uncertainty. 
    
    Further investigation is crucial to go beyond analytic forms of SFHs. We notice that non-parametric SFHs (e.g., \citealt{Leja2019ApJ}) could be more flexible options than these analytic approximations in describing the full diversity of SFH shapes, and they are becoming the recently popular alternatives to parametric ones. Therefore, we acknowledge that to assess systematic effects they could be interesting alternatives to exploit. We explored this possibility using the latest update of Bagpipes, which includes the possibility of using a non-parametric SFH proposed by \citep{Leja2019ApJ}, considering a model with a continuity prior on $\Delta\log(\mathrm{SFR})$ between adjacent time bins (7 bins in total). The results we obtained are extremely encouraging, and point toward the fact that the SFH uncertainty is currently not dominant in our analysis. We found that the Hubble parameter estimated considering non-parametric SFH is compatible within $0.27\sigma$ with respect to the one we derive in our analysis. A larger statistical uncertainty is also expected in this case, since a more flexible SFH with a larger number of free parameters is considered. However, further checks and verifications that go beyond the scope of this paper are needed before including this result as further systematic uncertainty in our analysis, and we defer it to a following paper.
    
    Dry mergers of less massive galaxies hosting younger stellar populations may bias the light-weighted age estimation toward younger ages. However, this effect is expected to be sub-dominant since this analysis is based on differential ages and spans a limited redshift interval \citep{Moresco:2012}. While the sample selection criteria in section~\ref{sec:data} are aimed to minimize contamination of the young component, a residual, even if minor, evolution could bias our measurement,  in \cite{Moresco2018ApJ} a recipe was provided to include in the covariance matrix of the $H(z)$ measurement an error contribution due to this effect, that we quantify to be negligible in this sample. 

    The rejuvenation in the star-formation history could introduce bias in the measurement of the differential ages. This has been addressed in detail in \cite{Moresco2012, Moresco2018ApJ}, where it has been quantified the impact of these effects on the Hubble parameter (see in particular App. A.1 of \cite{Moresco2012}, and Sect. 5 and 6 of \cite{Moresco2018ApJ}). The main point to stress here is that the selection process in section~\ref{sec:data}  and the differential approach of the cosmic chronometer method in section~\ref{sec:hubble_parameter} play a crucial role in minimizing this effect. Moreover, the systematic uncertainty originating from the binning method takes into account the potential progenitor bias, as we explain in section~\ref{sub:measuring_hz}.

    To include in the current measurement a proper full systematic covariance matrix, we therefore suggest the reader to follow the procedure described in details in \cite{Moresco2022} and \cite{Moresco2020ApJ}, including the missing statistical effects not already included in this analysis.

    In conclusion, this work provides a further important piece of evidence supporting the robustness of the CC method as a cosmological probe, showing the potential of the full-spectral fitting approach as another different method to derive the relative ages of massive and passive galaxies. It is interesting to notice here that our sample is almost a factor 2 larger than the final sample used by \cite{Borghi2022b}, and the statistical error between the two measurement scales as expected as $\sqrt{N}$. This is very promising in the view of several current \citep[SDSS BOSS Data Release 16,][]{boss16} and incoming spectroscopic surveys (such as Euclid \citealp{Laureijs2011} and \citealp{Atlasmission}), that will significantly improve the census of massive and passive galaxies, especially at $z>1$.

\begin{acknowledgments}
    We thank the anonymous referee for the constructive comments and suggestions, which help us greatly improve our manuscript. The first author is funded by the China Scholarship Council (CSC) from the Ministry of Education of P.R. China. This work is supported by the National Science Foundation of China(Grants No. 11929301) and the National Key R\&D Program of China(2017YFA0402600). N.B. and M.M. acknowledge support from MIUR, PRIN 2017 (grant 20179ZF5KS). M.M. acknowledges the grants ASI n.I/023/12/0 and ASI n.2018-23-HH.0. We acknowl- edge the use of computational resources from the parallel computing cluster of the Open Physics Hub (href{https://site.unibo.it/openphysicshub/en}{}) at the Physics and Astronomy Department in Bologna.
\end{acknowledgments}

\software{
        \bagpipes~ \citep{Carnall2018MNRAS},
        \textsc{MultiNest} \citep{2014BuchnerMULTINEST},
        \textsc{PyLick} \citep{Borghi2022a},
        \textsc{Astropy} \citep{Astropy2018},   
        \textsc{Matplotlib} \citep{Hunter2007},
        \textsc{Numpy} \citep{Harris2020},
        \scipy \citep{2020SciPy-NMeth}.
        }

\appendix
\section{The dust-metallicity degeneracy}\label{app:met}
\begin{figure}[h]
    \centering
        \includegraphics[width=.45\hsize]{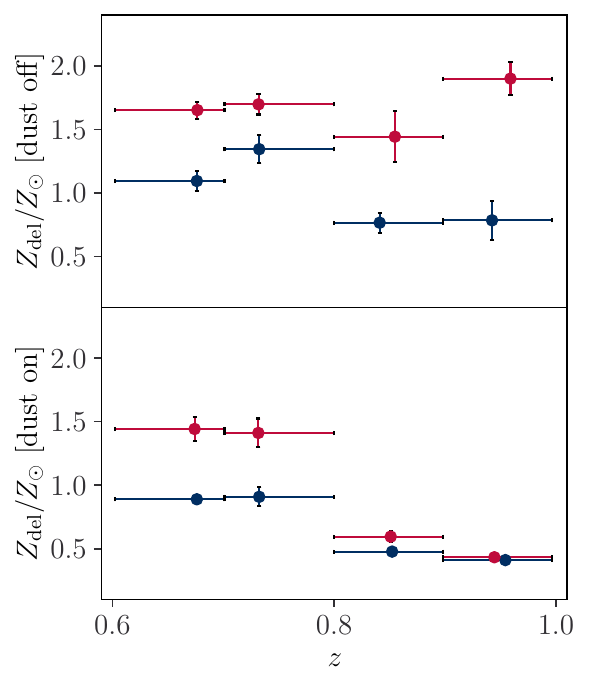}
        \caption{Median binned metallicity-redshift relations for our baseline results based on a full-spectrum fitting that contains a dust model (bottom, \citealt{Carnall2018MNRAS})  and for the fitting under the same configuration, but with the dust model switched off (upper). The blue and red points represent the lower and higher $\sigma_\star$ bins, divided using the median value of each sample as the threshold. For each bin, the vertical error bars are the errors associated to the median ages, 
        while horizontal bars denote the bin width.}\label{fig:dustonoff}
    \end{figure}

    In the posterior properties of the passive sample that is analyzed with the baseline configuration described in Section~\ref{sec:method}, we observe a small evolution of metallicity as a function of  redshift (as shown in the lower panel of Figure~\ref{fig:dustonoff}), which appears to contradict the pure passive selection assumption of our sample, contradicting the results by \citet{Borghi2022a} that on the same sample found no evident evolution in metallicity and stellar mass. We notice here that we observe no evolution in the stellar mass from the same baseline result. 
    
    We find that such evolution is due to a degeneracy between metallicity and dust in the fit. Since, as shown in the upper panel of Figure~\ref{fig:dustonoff}, the evolution completely disappears when we remove that parameter from the fit.  We find stellar metallicities with slightly supersolar values, $\left\langle Z/Z_\odot\right\rangle= 1.44\pm0.56$, in agreement with \citet{Borghi2022a} who find typical $Z/Z_\odot\sim 1.1$ using Lick indices. The metallicities show a clear downsizing pattern that, at each cosmic epoch, the stellar populations hosted in galaxies with higher mass are more metal-rich. In this configuration, we obtain $H(z=0.80) = 111.8 \pm 14.2~\mathrm{km~s^{-1}~Mpc^{-1}}$ based on the baseline configuration except for the dust model removed, supporting the differential age measured is very stable since the estimated Hubble parameter varies only by 1.1\% comparing to our baseline measurement, well below the currently estimated error. Since the main results are not affected, we decide to keep the current analysis as our baseline as it allows for a wider exploration of the parameter space.

\bibliography{main}{}

\begin{thebibliography}{}
\expandafter\ifx\csname natexlab\endcsname\relax\def\natexlab#1{#1}\fi
\providecommand{\url}[1]{\href{#1}{#1}}
\providecommand{\dodoi}[1]{doi:~\href{http://doi.org/#1}{\nolinkurl{#1}}}
\providecommand{\doeprint}[1]{\href{http://ascl.net/#1}{\nolinkurl{http://ascl.net/#1}}}
\providecommand{\doarXiv}[1]{\href{https://arxiv.org/abs/#1}{\nolinkurl{https://arxiv.org/abs/#1}}}

\bibitem[{{Abbott} {et~al.}(2017){Abbott}, {Abbott}, {Abbott}, {Acernese},
  {Ackley}, {Adams}, {Adams}, {Addesso}, {Adhikari}, {Adya}, {Affeldt},
  {Afrough}, {Agarwal}, {Agathos}, {Agatsuma}, {Aggarwal}, {Aguiar}, {Aiello},
  {Ain}, {Ajith}, {Allen}, {Allen}, {Allocca}, {Altin}, {Amato}, {Ananyeva},
  {Anderson}, {Anderson}, {Angelova}, {Antier}, {Appert}, {Arai}, {Araya},
  {Areeda}, {Arnaud}, {Arun}, {Ascenzi}, {Ashton}, {Ast}, {Aston}, {Astone},
  {Atallah}, {Aufmuth}, {Aulbert}, {Aultoneal}, {Austin}, {Avila-Alvarez},
  {Babak}, {Bacon}, {Bader}, {Bae}, {Baker}, {Baldaccini}, {Ballardin},
  {Ballmer}, {Banagiri}, {Barayoga}, {Barclay}, {Barish}, {Barker}, {Barkett},
  {Barone}, {Barr}, {Barsotti}, {Barsuglia}, {Barta}, {Bartlett}, {Bartos},
  {Bassiri}, {Basti}, {Batch}, {Bawaj}, {Bayley}, {Bazzan}, {B{\'e}csy},
  {Beer}, {Bejger}, {Belahcene}, {Bell}, {Berger}, {Bergmann}, {Bero}, {Berry},
  {Bersanetti}, {Bertolini}, {Betzwieser}, {Bhagwat}, {Bhandare}, {Bilenko},
  {Billingsley}, {Billman}, {Birch}, {Birney}, {Birnholtz}, {Biscans},
  {Biscoveanu}, {Bisht}, {Bitossi}, {Biwer}, {Bizouard}, {Blackburn},
  {Blackman}, {Blair}, {Blair}, {Blair}, {Bloemen}, {Bock}, {Bode}, {Boer},
  {Bogaert}, {Bohe}, {Bondu}, {Bonilla}, {Bonnand}, {Boom}, {Bork}, {Boschi},
  {Bose}, {Bossie}, {Bouffanais}, {Bozzi}, {Bradaschia}, {Brady}, {Branchesi},
  {Brau}, {Briant}, {Brillet}, {Brinkmann}, {Brisson}, {Brockill}, {Broida},
  {Brooks}, {Brown}, {Brown}, {Brunett}, {Buchanan}, {Buikema}, {Bulik},
  {Bulten}, {Buonanno}, {Buskulic}, {Buy}, {Byer}, {Cabero}, {Cadonati},
  {Cagnoli}, {Cahillane}, {Bustillo}, {Callister}, {Calloni}, {Camp}, {Canepa},
  {Canizares}, {Cannon}, {Cao}, {Cao}, {Capano}, {Capocasa}, {Carbognani},
  {Caride}, {Carney}, {Diaz}, {Casentini}, {Caudill}, {Cavagli{\`a}},
  {Cavalier}, {Cavalieri}, {Cella}, {Cepeda}, {Cerd{\'a}-Dur{\'a}n},
  {Cerretani}, {Cesarini}, {Chamberlin}, {Chan}, {Chao}, {Charlton}, {Chase},
  {Chassande-Mottin}, {Chatterjee}, {Chatziioannou}, {Cheeseboro}, {Chen},
  {Chen}, {Chen}, {Cheng}, {Chia}, {Chincarini}, {Chiummo}, {Chmiel}, {Cho},
  {Cho}, {Chow}, {Christensen}, {Chu}, {Chua}, {Chua}, {Chung}, {Chung},
  {Ciani}, {Ciolfi}, {Cirelli}, {Cirone}, {Clara}, {Clark}, {Clearwater},
  {Cleva}, {Cocchieri}, {Coccia}, {Cohadon}, {Cohen}, {Colla}, {Collette},
  {Cominsky}, {Constancio}, {Conti}, {Cooper}, {Corban}, {Corbitt},
  {Cordero-Carri{\'o}n}, {Corley}, {Cornish}, {Corsi}, {Cortese}, {Costa},
  {Coughlin}, {Coughlin}, {Coulon}, {Countryman}, {Couvares}, {Covas}, {Cowan},
  {Coward}, {Cowart}, {Coyne}, {Coyne}, {Creighton}, {Creighton}, {Cripe},
  {Crowder}, {Cullen}, {Cumming}, {Cunningham}, {Cuoco}, {Dal Canton},
  {D{\'a}lya}, {Danilishin}, {D'Antonio}, {Danzmann}, {Dasgupta}, {da Silva
  Costa}, {Datrier}, {Dattilo}, {Dave}, {Davier}, {Davis}, {Daw}, {Day}, {de},
  {Debra}, {Degallaix}, {de Laurentis}, {Del{\'e}glise}, {Del Pozzo}, {Demos},
  {Denker}, {Dent}, {de Pietri}, {Dergachev}, {De Rosa}, {Derosa}, {de Rossi},
  {Desalvo}, {de Varona}, {Devenson}, {Dhurandhar}, {D{\'\i}az}, {di Fiore},
  {di Giovanni}, {di Girolamo}, {di Lieto}, {di Pace}, {di Palma}, {di Renzo},
  {Doctor}, {Dolique}, {Donovan}, {Dooley}, {Doravari}, {Dorrington},
  {Douglas}, {Dovale {\'A}lvarez}, {Downes}, {Drago}, {Dreissigacker},
  {Driggers}, {Du}, {Ducrot}, {Dupej}, {Dwyer}, {Edo}, {Edwards}, {Effler},
  {Eggenstein}, {Ehrens}, {Eichholz}, {Eikenberry}, {Eisenstein}, {Essick},
  {Estevez}, {Etienne}, {Etzel}, {Evans}, {Evans}, {Factourovich}, {Fafone},
  {Fair}, {Fairhurst}, {Fan}, {Farinon}, {Farr}, {Farr}, {Fauchon-Jones},
  {Favata}, {Fays}, {Fee}, {Fehrmann}, {Feicht}, {Fejer}, {Fernandez-Galiana},
  {Ferrante}, {Ferreira}, {Ferrini}, {Fidecaro}, {Finstad}, {Fiori},
  {Fiorucci}, {Fishbach}, {Fisher}, {Fitz-Axen}, {Flaminio}, {Fletcher},
  {Fong}, {Font}, {Forsyth}, {Forsyth}, {Fournier}, {Frasca}, {Frasconi},
  {Frei}, {Freise}, {Frey}, {Frey}, {Fries}, {Fritschel}, {Frolov}, {Fulda},
  {Fyffe}, {Gabbard}, {Gadre}, {Gaebel}, {Gair}, {Gammaitoni}, {Ganija},
  {Gaonkar}, {Garcia-Quiros}, {Garufi}, {Gateley}, {Gaudio}, {Gaur},
  {Gayathri}, {Gehrels}, {Gemme}, {Genin}, {Gennai}, {George}, {George},
  {Gergely}, {Germain}, {Ghonge}, {Ghosh}, {Ghosh}, {Ghosh}, {Giaime},
  {Giardina}, {Giazotto}, {Gill}, {Glover}, {Goetz}, {Goetz}, {Gomes},
  {Goncharov}, {Gonz{\'a}lez}, {Castro}, {Gopakumar}, {Gorodetsky}, {Gossan},
  {Gosselin}, {Gouaty}, {Grado}, {Graef}, {Granata}, {Grant}, {Gras}, {Gray},
  {Greco}, {Green}, {Gretarsson}, {Groot}, {Grote}, {Grunewald}, {Gruning},
  {Guidi}, {Guo}, {Gupta}, {Gupta}, {Gushwa}, {Gustafson}, {Gustafson},
  {Halim}, {Hall}, {Hall}, {Hamilton}, {Hammond}, {Haney}, {Hanke}, {Hanks},
  {Hanna}, {Hannam}, {Hannuksela}, {Hanson}, {Hardwick}, {Harms}, {Harry},
  {Harry}, {Hart}, {Haster}, {Haughian}, {Healy}, {Heidmann}, {Heintze},
  {Heitmann}, {Hello}, {Hemming}, {Hendry}, {Heng}, {Hennig}, {Heptonstall},
  {Heurs}, {Hild}, {Hinderer}, {Hoak}, {Hofman}, {Holt}, {Holz}, {Hopkins},
  {Horst}, {Hough}, {Houston}, {Howell}, {Hreibi}, {Hu}, {Huerta}, {Huet},
  {Hughey}, {Husa}, {Huttner}, {Huynh-Dinh}, {Indik}, {Inta}, {Intini}, {Isa},
  {Isac}, {Isi}, {Iyer}, {Izumi}, {Jacqmin}, {Jani}, {Jaranowski}, {Jawahar},
  {Jim{\'e}nez-Forteza}, {Johnson}, {Jones}, {Jones}, {Jonker}, {Ju}, {Junker},
  {Kalaghatgi}, {Kalogera}, {Kamai}, {Kandhasamy}, {Kang}, {Kanner}, {Kapadia},
  {Karki}, {Karvinen}, {Kasprzack}, {Katolik}, {Katsavounidis}, {Katzman},
  {Kaufer}, {Kawabe}, {K{\'e}f{\'e}lian}, {Keitel}, {Kemball}, {Kennedy},
  {Kent}, {Key}, {Khalili}, {Khan}, {Khan}, {Khan}, {Khazanov}, {Kijbunchoo},
  {Kim}, {Kim}, {Kim}, {Kim}, {Kim}, {Kim}, {Kimbrell}, {King}, {King},
  {Kinley-Hanlon}, {Kirchhoff}, {Kissel}, {Kleybolte}, {Klimenko}, {Knowles},
  {Koch}, {Koehlenbeck}, {Koley}, {Kondrashov}, {Kontos}, {Korobko}, {Korth},
  {Kowalska}, {Kozak}, {Kr{\"a}mer}, {Kringel}, {Krishnan}, {Kr{\'o}lak},
  {Kuehn}, {Kumar}, {Kumar}, {Kumar}, {Kuo}, {Kutynia}, {Kwang}, {Lackey},
  {Lai}, {Landry}, {Lang}, {Lange}, {Lantz}, {Lanza}, {Lartaux-Vollard},
  {Lasky}, {Laxen}, {Lazzarini}, {Lazzaro}, {Leaci}, {Leavey}, {Lee}, {Lee},
  {Lee}, {Lee}, {Lee}, {Lehmann}, {Lenon}, {Leonardi}, {Leroy}, {Letendre},
  {Levin}, {Li}, {Linker}, {Littenberg}, {Liu}, {Liu}, {Lo}, {Lockerbie},
  {London}, {Lord}, {Lorenzini}, {Loriette}, {Lormand}, {Losurdo}, {Lough},
  {Lousto}, {Lovelace}, {L{\"u}ck}, {Lumaca}, {Lundgren}, {Lynch}, {Ma},
  {Macas}, {Macfoy}, {Machenschalk}, {Macinnis}, {MacLeod}, {Hernandez},
  {Maga{\~n}a-Sandoval}, {Zertuche}, {Magee}, {Majorana}, {Maksimovic}, {Man},
  {Mandic}, {Mangano}, {Mansell}, {Manske}, {Mantovani}, {Marchesoni},
  {Marion}, {M{\'a}rka}, {M{\'a}rka}, {Markakis}, {Markosyan}, {Markowitz},
  {Maros}, {Marquina}, {Martelli}, {Martellini}, {Martin}, {Martin},
  {Martynov}, {Mason}, {Massera}, {Masserot}, {Massinger}, {Masso-Reid},
  {Mastrogiovanni}, {Matas}, {Matichard}, {Matone}, {Mavalvala}, {Mazumder},
  {McCarthy}, {McClelland}, {McCormick}, {McCuller}, {McGuire}, {McIntyre},
  {McIver}, {McManus}, {McNeill}, {McRae}, {McWilliams}, {Meacher}, {Meadors},
  {Mehmet}, {Meidam}, {Mejuto-Villa}, {Melatos}, {Mendell}, {Mercer}, {Merilh},
  {Merzougui}, {Meshkov}, {Messenger}, {Messick}, {Metzdorff}, {Meyers},
  {Miao}, {Michel}, {Middleton}, {Mikhailov}, {Milano}, {Miller}, {Miller},
  {Miller}, {Millhouse}, {Milovich-Goff}, {Minazzoli}, {Minenkov}, {Ming},
  {Mishra}, {Mitra}, {Mitrofanov}, {Mitselmakher}, {Mittleman}, {Moffa},
  {Moggi}, {Mogushi}, {Mohan}, {Mohapatra}, {Montani}, {Moore}, {Moraru},
  {Moreno}, {Morriss}, {Mours}, {Mow-Lowry}, {Mueller}, {Muir}, {Mukherjee},
  {Mukherjee}, {Mukherjee}, {Mukund}, {Mullavey}, {Munch}, {Mu{\~n}iz},
  {Muratore}, {Murray}, {Napier}, {Nardecchia}, {Naticchioni}, {Nayak},
  {Neilson}, {Nelemans}, {Nelson}, {Nery}, {Neunzert}, {Nevin}, {Newport},
  {Newton}, {Ng}, {Nguyen}, {Nichols}, {Nielsen}, {Nissanke}, {Nitz}, {Noack},
  {Nocera}, {Nolting}, {North}, {Nuttall}, {Oberling}, {O'Dea}, {Ogin}, {Oh},
  {Oh}, {Ohme}, {Okada}, {Oliver}, {Oppermann}, {Oram}, {O'Reilly}, {Ormiston},
  {Ortega}, {O'Shaughnessy}, {Ossokine}, {Ottaway}, {Overmier}, {Owen}, {Pace},
  {Page}, {Page}, {Pai}, {Pai}, {Palamos}, {Palashov}, {Palomba}, {Pal-Singh},
  {Pan}, {Pan}, {Pang}, {Pang}, {Pankow}, {Pannarale}, {Pant}, {Paoletti},
  {Paoli}, {Papa}, {Parida}, {Parker}, {Pascucci}, {Pasqualetti},
  {Passaquieti}, {Passuello}, {Patil}, {Patricelli}, {Pearlstone}, {Pedraza},
  {Pedurand}, {Pekowsky}, {Pele}, {Penn}, {Perez}, {Perreca}, {Perri},
  {Pfeiffer}, {Phelps}, {Piccinni}, {Pichot}, {Piergiovanni}, {Pierro},
  {Pillant}, {Pinard}, {Pinto}, {Pirello}, {Pitkin}, {Poe}, {Poggiani},
  {Popolizio}, {Porter}, {Post}, {Powell}, {Prasad}, {Pratt}, {Pratten},
  {Predoi}, {Prestegard}, {Prijatelj}, {Principe}, {Privitera}, {Prodi},
  {Prokhorov}, {Puncken}, {Punturo}, {Puppo}, {P{\"u}rrer}, {Qi}, {Quetschke},
  {Quintero}, {Quitzow-James}, {Raab}, {Rabeling}, {Radkins}, {Raffai}, {Raja},
  {Rajan}, {Rajbhandari}, {Rakhmanov}, {Ramirez}, {Ramos-Buades}, {Rapagnani},
  {Raymond}, {Razzano}, {Read}, {Regimbau}, {Rei}, {Reid}, {Reitze}, {Ren},
  {Reyes}, {Ricci}, {Ricker}, {Rieger}, {Riles}, {Rizzo}, {Robertson}, {Robie},
  {Robinet}, {Rocchi}, {Rolland}, {Rollins}, {Roma}, {Romano}, {Romano},
  {Romel}, {Romie}, {Rosi{\'n}ska}, {Ross}, {Rowan}, {R{\"u}diger}, {Ruggi},
  {Rutins}, {Ryan}, {Sachdev}, {Sadecki}, {Sadeghian}, {Sakellariadou},
  {Salconi}, {Saleem}, {Salemi}, {Samajdar}, {Sammut}, {Sampson}, {Sanchez},
  {Sanchez}, {Sanchis-Gual}, {Sandberg}, {Sanders}, {Sassolas},
  {Sathyaprakash}, {Saulson}, {Sauter}, {Savage}, {Sawadsky}, {Schale},
  {Scheel}, {Scheuer}, {Schmidt}, {Schmidt}, {Schnabel}, {Schofield},
  {Sch{\"o}nbeck}, {Schreiber}, {Schuette}, {Schulte}, {Schutz}, {Schwalbe},
  {Scott}, {Scott}, {Seidel}, {Sellers}, {Sengupta}, {Sentenac}, {Sequino},
  {Sergeev}, {Shaddock}, {Shaffer}, {Shah}, {Shahriar}, {Shaner}, {Shao},
  {Shapiro}, {Shawhan}, {Sheperd}, {Shoemaker}, {Shoemaker}, {Siellez},
  {Siemens}, {Sieniawska}, {Sigg}, {Silva}, {Singer}, {Singh}, {Singhal},
  {Sintes}, {Slagmolen}, {Smith}, {Smith}, {Smith}, {Somala}, {Son},
  {Sonnenberg}, {Sorazu}, {Sorrentino}, {Souradeep}, {Spencer}, {Srivastava},
  {Staats}, {Staley}, {Steer}, {Steinke}, {Steinlechner}, {Steinlechner},
  {Steinmeyer}, {Stevenson}, {Stone}, {Stops}, {Strain}, {Stratta}, {Strigin},
  {Strunk}, {Sturani}, {Stuver}, {Summerscales}, {Sun}, {Sunil}, {Suresh},
  {Sutton}, {Swinkels}, {Szczepa{\'n}czyk}, {Tacca}, {Tait}, {Talbot},
  {Talukder}, {Tanner}, {T{\'a}pai}, {Taracchini}, {Tasson}, {Taylor},
  {Taylor}, {Tewari}, {Theeg}, {Thies}, {Thomas}, {Thomas}, {Thomas}, {Thorne},
  {Thrane}, {Tiwari}, {Tiwari}, {Tokmakov}, {Toland}, {Tonelli}, {Tornasi},
  {Torres-Forn{\'e}}, {Torrie}, {T{\"o}yr{\"a}}, {Travasso}, {Traylor},
  {Trinastic}, {Tringali}, {Trozzo}, {Tsang}, {Tse}, {Tso}, {Tsukada}, {Tsuna},
  {Tuyenbayev}, {Ueno}, {Ugolini}, {Unnikrishnan}, {Urban}, {Usman},
  {Vahlbruch}, {Vajente}, {Valdes}, {van Bakel}, {van Beuzekom}, {van den
  Brand}, {van den Broeck}, {Vander-Hyde}, {van der Schaaf}, {van Heijningen},
  {van Veggel}, {Vardaro}, {Varma}, {Vass}, {Vas{\'u}th}, {Vecchio},
  {Vedovato}, {Veitch}, {Veitch}, {Venkateswara}, {Venugopalan}, {Verkindt},
  {Vetrano}, {Vicer{\'e}}, {Viets}, {Vinciguerra}, {Vine}, {Vinet}, {Vitale},
  {Vo}, {Vocca}, {Vorvick}, {Vyatchanin}, {Wade}, {Wade}, {Wade}, {Walet},
  {Walker}, {Wallace}, {Walsh}, {Wang}, {Wang}, {Wang}, {Wang}, {Wang}, {Ward},
  {Warner}, {Was}, {Watchi}, {Weaver}, {Wei}, {Weinert}, {Weinstein}, {Weiss},
  {Wen}, {Wessel}, {We{\ss}els}, {Westerweck}, {Westphal}, {Wette}, {Whelan},
  {Whitcomb}, {Whiting}, {Whittle}, {Wilken}, {Williams}, {Williams},
  {Williamson}, {Willis}, {Willke}, {Wimmer}, {Winkler}, {Wipf}, {Wittel},
  {Woan}, {Woehler}, {Wofford}, {Wong}, {Worden}, {Wright}, {Wu}, {Wysocki},
  {Xiao}, {Yamamoto}, {Yancey}, {Yang}, {Yap}, {Yazback}, {Yu}, {Yu}, {Yvert},
  {Zadro{\.z}ny}, {Zanolin}, {Zelenova}, {Zendri}, {Zevin}, {Zhang}, {Zhang},
  {Zhang}, {Zhang}, {Zhao}, {Zhou}, {Zhou}, {Zhu}, {Zhu}, {Zimmerman},
  {Zucker}, {Zweizig}, {Foley}, {Coulter}, {Drout}, {Kasen}, {Kilpatrick},
  {Madore}, {Murguia-Berthier}, {Pan}, {Piro}, {Prochaska}, {Ramirez-Ruiz},
  {Rest}, {Rojas-Bravo}, {Shappee}, {Siebert}, {Simon}, {Ulloa}, {Annis},
  {Soares-Santos}, {Brout}, {Scolnic}, {Diehl}, {Frieman}, {Berger},
  {Alexander}, {Allam}, {Balbinot}, {Blanchard}, {Butler}, {Chornock}, {Cook},
  {Cowperthwaite}, {Drlica-Wagner}, {Drout}, {Durret}, {Eftekhari}, {Finley},
  {Fong}, {Fryer}, {Garc{\'\i}a-Bellido}, {Gill}, {Gruendl}, {Hanna},
  {Hartley}, {Herner}, {Huterer}, {Kasen}, {Kessler}, {Li}, {Lin}, {Lopes},
  {Louren{\c{c}}o}, {Margutti}, {Marriner}, {Marshall}, {Matheson}, {Medina},
  {Metzger}, {Mu{\~n}oz}, {Muir}, {Nicholl}, {Nugent}, {Palmese},
  {Paz-Chinch{\'o}n}, {Quataert}, {Sako}, {Sauseda}, {Schlegel}, {Secco},
  {Smith}, {Sobreira}, {Stebbins}, {Villar}, {Vivas}, {Wester}, {Williams},
  {Yanny}, {Zenteno}, {Abbott}, {Abdalla}, {Bechtol}, {Benoit-L{\'e}vy},
  {Bertin}, {Bridle}, {Brooks}, {Buckley-Geer}, {Burke}, {Rosell}, {Kind},
  {Carretero}, {Castander}, {Cunha}, {D'Andrea}, {da Costa}, {Davis}, {Depoy},
  {Desai}, {Dietrich}, {Estrada}, {Fernandez}, {Flaugher}, {Fosalba},
  {Gaztanaga}, {Gerdes}, {Giannantonio}, {Goldstein}, {Gruen}, {Gutierrez},
  {Hartley}, {Honscheid}, {Jain}, {James}, {Jeltema}, {Johnson}, {Kent},
  {Krause}, {Kron}, {Kuehn}, {Kuhlmann}, {Kuropatkin}, {Lahav}, {Lima}, {Maia},
  {March}, {Miller}, {Miquel}, {Neilsen}, {Nord}, {Ogando}, {Plazas}, {Romer},
  {Roodman}, {Rykoff}, {Sanchez}, {Scarpine}, {Schubnell}, {Sevilla-Noarbe},
  {Smith}, {Smith}, {Suchyta}, {Tarle}, {Thomas}, {Thomas}, {Troxel}, {Tucker},
  {Vikram}, {Walker}, {Weller}, {Zhang}, {Haislip}, {Kouprianov}, {Reichart},
  {Tartaglia}, {Sand}, {Valenti}, {Yang}, {Arcavi}, {Hosseinzadeh}, {Howell},
  {McCully}, {Poznanski}, {Vasylyev}, {Tanvir}, {Levan}, {Hjorth}, {Cano},
  {Copperwheat}, {de Ugarte-Postigo}, {Evans}, {Fynbo},
  {Gonz{\'a}lez-Fern{\'a}ndez}, {Greiner}, {Irwin}, {Lyman}, {Mandel},
  {McMahon}, {Milvang-Jensen}, {O'Brien}, {Osborne}, {Perley}, {Pian},
  {Palazzi}, {Rol}, {Rosetti}, {Rosswog}, {Rowlinson}, {Schulze}, {Steeghs},
  {Th{\"o}ne}, {Ulaczyk}, {Watson}, {Wiersema}, {Lipunov}, {Gorbovskoy},
  {Kornilov}, {Tyurina}, {Balanutsa}, {Vlasenko}, {Gorbunov}, {Podesta},
  {Levato}, {Saffe}, {Buckley}, {Budnev}, {Gress}, {Yurkov}, {Rebolo}, \&
  {Serra-Ricart}}]{Abbott2017Natur}
{Abbott}, B.~P., {Abbott}, R., {Abbott}, T.~D., {et~al.} 2017, \nat, 551, 85,
  \dodoi{10.1038/nature24471}

\bibitem[{{Abdalla} {et~al.}(2022){Abdalla}, {Abell{\'a}n}, {Aboubrahim},
  {Agnello}, {Akarsu}, {Akrami}, {Alestas}, {Aloni}, {Amendola}, {Anchordoqui},
  {Anderson}, {Arendse}, {Asgari}, {Ballardini}, {Barger}, {Basilakos},
  {Batista}, {Battistelli}, {Battye}, {Benetti}, {Benisty}, {Berlin}, {de
  Bernardis}, {Berti}, {Bidenko}, {Birrer}, {Blakeslee}, {Boddy}, {Bom},
  {Bonilla}, {Borghi}, {Bouchet}, {Braglia}, {Buchert}, {Buckley-Geer},
  {Calabrese}, {Caldwell}, {Camarena}, {Capozziello}, {Casertano}, {Chen},
  {Chluba}, {Chen}, {Chen}, {Chudaykin}, {Cicoli}, {Copi}, {Courbin},
  {Cyr-Racine}, {Czerny}, {Dainotti}, {D'Amico}, {Davis}, {de Cruz P{\'e}rez},
  {de Haro}, {Delabrouille}, {Denton}, {Dhawan}, {Dienes}, {Di Valentino},
  {Du}, {Eckert}, {Escamilla-Rivera}, {Fert{\'e}}, {Finelli}, {Fosalba},
  {Freedman}, {Frusciante}, {Gazta{\~n}aga}, {Giar{\`e}}, {Giusarma},
  {G{\'o}mez-Valent}, {Handley}, {Harrison}, {Hart}, {Hazra}, {Heavens},
  {Heinesen}, {Hildebrandt}, {Hill}, {Hogg}, {Holz}, {Hooper}, {Hosseininejad},
  {Huterer}, {Ishak}, {Ivanov}, {Jaffe}, {Jang}, {Jedamzik}, {Jimenez},
  {Joseph}, {Joudaki}, {Kamionkowski}, {Karwal}, {Kazantzidis}, {Keeley},
  {Klasen}, {Komatsu}, {Koopmans}, {Kumar}, {Lamagna}, {Lazkoz}, {Lee},
  {Lesgourgues}, {Levi Said}, {Lewis}, {L'Huiller}, {Lucca}, {Maartens},
  {Macri}, {Marfatia}, {Marra}, {Martins}, {Masi}, {Matarrese}, {Mazumdar},
  {Melchiorri}, {Mena}, {Mersini-Houghton}, {Mertens}, {Milakovic}, {Minami},
  {Miranda}, {Moreno-Pulido}, {Moresco}, {Mota}, {Mottola}, {Mozzon}, {Muir},
  {Mukherjee}, {Mukherjee}, {Naselsky}, {Nath}, {Nesseris}, {Niedermann},
  {Notari}, {Nunes}, {Colg{\'a}in}, {Owens}, {Ozulker}, {Pace},
  {Paliathanasis}, {Palmese}, {Pan}, {Paoletti}, {Perez Bergliaffa},
  {Perivolaropoulos}, {Pesce}, {Pettorino}, {Philcox}, {Pogosian}, {Poulin},
  {Poulot}, {Raveri}, {Reid}, {Renzi}, {Riess}, {Sabla}, {Salucci}, {Salzano},
  {Saridakis}, {Sathyaprakash}, {Schmaltz}, {Sch{\"o}neberg}, {Scolnic}, {Sen},
  {Sehgal}, {Shafieloo}, {Sheikh-Jabbari}, {Silk}, {Silvestri}, {Skara},
  {Sloth}, {Soares-Santos}, {Sol{\`a} Peracaula}, {Songsheng}, {Soriano},
  {Staicova}, {Starkman}, {Szapudi}, {Teixera}, {Thomas}, {Treu}, {Trott}, {van
  de Bruck}, {Vazquez}, {Verde}, {Visinelli}, {Wang}, {Wang}, {Wang},
  {Watkins}, {Watson}, {Webb}, {Weiner}, {Weltman}, {Witte}, {Wojtak}, {Yadav},
  {Yang}, {Zhao}, \& {Zumalac{\'a}rregui}}]{Abdalla2022}
{Abdalla}, E., {Abell{\'a}n}, G.~F., {Aboubrahim}, A., {et~al.} 2022, arXiv
  e-prints, arXiv:2203.06142.
\newblock \doarXiv{2203.06142}

\bibitem[{{Ahumada} {et~al.}(2020){Ahumada}, {Prieto}, {Almeida}, {Anders},
  {Anderson}, {Andrews}, {Anguiano}, {Arcodia}, {Armengaud}, {Aubert}, {Avila},
  {Avila-Reese}, {Badenes}, {Balland}, \& et~al.}]{boss16}
{Ahumada}, R., {Prieto}, C.~A., {Almeida}, A., {et~al.} 2020, \apjs, 249, 3,
  \dodoi{10.3847/1538-4365/ab929e}

\bibitem[{{Allen} {et~al.}(2011){Allen}, {Evrard}, \&
  {Mantz}}]{Allen2011ARA&A..49..409A}
{Allen}, S.~W., {Evrard}, A.~E., \& {Mantz}, A.~B. 2011, \araa, 49, 409,
  \dodoi{10.1146/annurev-astro-081710-102514}

\bibitem[{{Astropy Collaboration} {et~al.}(2018){Astropy Collaboration},
  {Price-Whelan}, {Sip{\H{o}}cz}, {G{\"u}nther}, {Lim}, {Crawford}, {Conseil},
  {Shupe}, {Craig}, {Dencheva}, {Ginsburg}, {VanderPlas}, {Bradley},
  {P{\'e}rez-Su{\'a}rez}, {de Val-Borro}, {Aldcroft}, {Cruz}, {Robitaille},
  {Tollerud}, {Ardelean}, {Babej}, {Bach}, {Bachetti}, {Bakanov}, {Bamford},
  {Barentsen}, {Barmby}, {Baumbach}, {Berry}, {Biscani}, {Boquien}, {Bostroem},
  {Bouma}, {Brammer}, {Bray}, {Breytenbach}, {Buddelmeijer}, {Burke},
  {Calderone}, {Cano Rodr{\'\i}guez}, {Cara}, {Cardoso}, {Cheedella}, {Copin},
  {Corrales}, {Crichton}, {D'Avella}, {Deil}, {Depagne}, {Dietrich}, {Donath},
  {Droettboom}, {Earl}, {Erben}, {Fabbro}, {Ferreira}, {Finethy}, {Fox},
  {Garrison}, {Gibbons}, {Goldstein}, {Gommers}, {Greco}, {Greenfield},
  {Groener}, {Grollier}, {Hagen}, {Hirst}, {Homeier}, {Horton}, {Hosseinzadeh},
  {Hu}, {Hunkeler}, {Ivezi{\'c}}, {Jain}, {Jenness}, {Kanarek}, {Kendrew},
  {Kern}, {Kerzendorf}, {Khvalko}, {King}, {Kirkby}, {Kulkarni}, {Kumar},
  {Lee}, {Lenz}, {Littlefair}, {Ma}, {Macleod}, {Mastropietro}, {McCully},
  {Montagnac}, {Morris}, {Mueller}, {Mumford}, {Muna}, {Murphy}, {Nelson},
  {Nguyen}, {Ninan}, {N{\"o}the}, {Ogaz}, {Oh}, {Parejko}, {Parley}, {Pascual},
  {Patil}, {Patil}, {Plunkett}, {Prochaska}, {Rastogi}, {Reddy Janga},
  {Sabater}, {Sakurikar}, {Seifert}, {Sherbert}, {Sherwood-Taylor}, {Shih},
  {Sick}, {Silbiger}, {Singanamalla}, {Singer}, {Sladen}, {Sooley},
  {Sornarajah}, {Streicher}, {Teuben}, {Thomas}, {Tremblay}, {Turner},
  {Terr{\'o}n}, {van Kerkwijk}, {de la Vega}, {Watkins}, {Weaver}, {Whitmore},
  {Woillez}, {Zabalza}, \& {Astropy Contributors}}]{Astropy2018}
{Astropy Collaboration}, {Price-Whelan}, A.~M., {Sip{\H{o}}cz}, B.~M., {et~al.}
  2018, \aj, 156, 123, \dodoi{10.3847/1538-3881/aabc4f}

\bibitem[{{Bartelmann} \& {Schneider}(2001)}]{Bartelmann2001PhR...340..291B}
{Bartelmann}, M., \& {Schneider}, P. 2001, \physrep, 340, 291,
  \dodoi{10.1016/S0370-1573(00)00082-X}

\bibitem[{Belli {et~al.}(2019)Belli, Newman, \& Ellis}]{Belli2019}
Belli, S., Newman, A.~B., \& Ellis, R.~S. 2019, \apj, 874, 17,
  \dodoi{10.3847/1538-4357/ab07af}

\bibitem[{{Ben{\'\i}tez} {et~al.}(2009){Ben{\'\i}tez}, {Gazta{\~n}aga},
  {Miquel}, {Castander}, {Moles}, {Crocce}, {Fern{\'a}ndez-Soto}, {Fosalba},
  {Ballesteros}, {Campa}, {Cardiel-Sas}, {Castilla}, {Crist{\'o}bal-Hornillos},
  {Delfino}, {Fern{\'a}ndez}, {Fern{\'a}ndez-Sopuerta}, {Garc{\'\i}a-Bellido},
  {Lobo}, {Mart{\'\i}nez}, {Ortiz}, {Pacheco}, {Paredes}, {Pons-Border{\'\i}a},
  {S{\'a}nchez}, {S{\'a}nchez}, {Varela}, \& {de Vicente}}]{benitez2009}
{Ben{\'\i}tez}, N., {Gazta{\~n}aga}, E., {Miquel}, R., {et~al.} 2009, \apj,
  691, 241, \dodoi{10.1088/0004-637X/691/1/241}

\bibitem[{{Bennett} {et~al.}(2003){Bennett}, {Halpern}, {Hinshaw}, {Jarosik},
  {Kogut}, {Limon}, {Meyer}, {Page}, {Spergel}, {Tucker}, {Wollack}, {Wright},
  {Barnes}, {Greason}, {Hill}, {Komatsu}, {Nolta}, {Odegard}, {Peiris},
  {Verde}, \& {Weiland}}]{Bennett2003ApJS}
{Bennett}, C.~L., {Halpern}, M., {Hinshaw}, G., {et~al.} 2003, \apjs, 148, 1,
  \dodoi{10.1086/377253}

\bibitem[{{Betoule} {et~al.}(2014){Betoule}, {Kessler}, {Guy}, {Mosher},
  {Hardin}, {Biswas}, {Astier}, {El-Hage}, {Konig}, {Kuhlmann}, {Marriner},
  {Pain}, {Regnault}, {Balland}, {Bassett}, {Brown}, {Campbell}, {Carlberg},
  {Cellier-Holzem}, {Cinabro}, {Conley}, {D'Andrea}, {DePoy}, {Doi}, {Ellis},
  {Fabbro}, {Filippenko}, {Foley}, {Frieman}, {Fouchez}, {Galbany}, {Goobar},
  {Gupta}, {Hill}, {Hlozek}, {Hogan}, {Hook}, {Howell}, {Jha}, {Le Guillou},
  {Leloudas}, {Lidman}, {Marshall}, {M{\"o}ller}, {Mour{\~a}o}, {Neveu},
  {Nichol}, {Olmstead}, {Palanque-Delabrouille}, {Perlmutter}, {Prieto},
  {Pritchet}, {Richmond}, {Riess}, {Ruhlmann-Kleider}, {Sako}, {Schahmaneche},
  {Schneider}, {Smith}, {Sollerman}, {Sullivan}, {Walton}, \&
  {Wheeler}}]{Betoule2014AA}
{Betoule}, M., {Kessler}, R., {Guy}, J., {et~al.} 2014, \aap, 568, A22,
  \dodoi{10.1051/0004-6361/201423413}

\bibitem[{Beverage {et~al.}(2021)Beverage, Kriek, Conroy, Bezanson, Franx, \&
  van~der Wel}]{Beverage2021}
Beverage, A.~G., Kriek, M., Conroy, C., {et~al.} 2021, \apjl, 917, L1,
  \dodoi{10.3847/2041-8213/ac12cd}

\bibitem[{Borghi {et~al.}(2022b)Borghi, Moresco, \& Cimatti}]{Borghi2022b}
Borghi, N., Moresco, M., \& Cimatti, A. 2022b, \apjl, 928, L4,
  \dodoi{10.3847/2041-8213/ac3fb2}

\bibitem[{Borghi {et~al.}(2022a)Borghi, Moresco, Cimatti, Huchet, Quai, \&
  Pozzetti}]{Borghi2022a}
Borghi, N., Moresco, M., Cimatti, A., {et~al.} 2022a, \apj, 927, 164,
  \dodoi{10.3847/1538-4357/ac3240}

\bibitem[{{Bruzual} \& {Charlot}(2003)}]{BC2003MNRAS}
{Bruzual}, G., \& {Charlot}, S. 2003, \mnras, 344, 1000,
  \dodoi{10.1046/j.1365-8711.2003.06897.x}

\bibitem[{{Buchner} {et~al.}(2014){Buchner}, {Georgakakis}, {Nandra}, \&
  et~al.}]{2014BuchnerMULTINEST}
{Buchner}, J., {Georgakakis}, A., {Nandra}, K., \& et~al. 2014, \aap, 564,
  A125, \dodoi{10.1051/0004-6361/201322971}

\bibitem[{{Calzetti} {et~al.}(2000){Calzetti}, {Armus}, {Bohlin}, {Kinney},
  {Koornneef}, \& {Storchi-Bergmann}}]{Calzetti2000ApJ}
{Calzetti}, D., {Armus}, L., {Bohlin}, R.~C., {et~al.} 2000, \apj, 533, 682,
  \dodoi{10.1086/308692}

\bibitem[{{Carlstrom} {et~al.}(2011){Carlstrom}, {Ade}, {Aird}, {Benson},
  {Bleem}, {Busetti}, {Chang}, {Chauvin}, {Cho}, {Crawford}, {Crites}, {Dobbs},
  {Halverson}, {Heimsath}, {Holzapfel}, {Hrubes}, {Joy}, {Keisler}, {Lanting},
  {Lee}, {Leitch}, {Leong}, {Lu}, {Lueker}, {Luong-Van}, {McMahon}, {Mehl},
  {Meyer}, {Mohr}, {Montroy}, {Padin}, {Plagge}, {Pryke}, {Ruhl}, {Schaffer},
  {Schwan}, {Shirokoff}, {Spieler}, {Staniszewski}, {Stark}, {Tucker},
  {Vanderlinde}, {Vieira}, \& {Williamson}}]{Carlstrom2011PASP}
{Carlstrom}, J.~E., {Ade}, P.~A.~R., {Aird}, K.~A., {et~al.} 2011, \pasp, 123,
  568, \dodoi{10.1086/659879}

\bibitem[{{Carnall} {et~al.}(2018){Carnall}, {McLure}, {Dunlop}, \&
  {Dav{\'e}}}]{Carnall2018MNRAS}
{Carnall}, A.~C., {McLure}, R.~J., {Dunlop}, J.~S., \& {Dav{\'e}}, R. 2018,
  \mnras, 480, 4379, \dodoi{10.1093/mnras/sty2169}

\bibitem[{{Carnall} {et~al.}(2019){Carnall}, {McLure}, {Dunlop}, {Cullen},
  {McLeod}, {Wild}, {Johnson}, {Appleby}, {Dav{\'e}}, {Amorin}, {Bolzonella},
  {Castellano}, {Cimatti}, {Cucciati}, {Gargiulo}, {Garilli}, {Marchi},
  {Pentericci}, {Pozzetti}, {Schreiber}, {Talia}, \&
  {Zamorani}}]{Carnall2019MNRAS}
{Carnall}, A.~C., {McLure}, R.~J., {Dunlop}, J.~S., {et~al.} 2019, \mnras, 490,
  417, \dodoi{10.1093/mnras/stz2544}

\bibitem[{Carnall {et~al.}(2022)Carnall, McLure, Dunlop, Hamadouche, Cullen,
  McLeod, Begley, Amorin, Bolzonella, Castellano, Cimatti, Fontanot, Gargiulo,
  Garilli, Mannucci, Pentericci, Talia, Zamorani, Calabro, Cresci, \&
  Hathi}]{Carnall2022}
Carnall, A.~C., McLure, R.~J., Dunlop, J.~S., {et~al.} 2022, \apj, 929, 131,
  \dodoi{10.3847/1538-4357/ac5b62}

\bibitem[{{Charlot} \& {Fall}(2000)}]{Charlot2000ApJ}
{Charlot}, S., \& {Fall}, S.~M. 2000, \apj, 539, 718, \dodoi{10.1086/309250}

\bibitem[{Chauke {et~al.}(2018)Chauke, van~der Wel, Pacifici, Bezanson, Wu,
  Gallazzi, Noeske, Straatman, Mu{\~{n}}os-Mateos, Franx, Bari{\v{s}}i{\'{c}},
  Bell, Brammer, Calhau, van Houdt, Labb{\'{e}}, Maseda, Muzzin, Rix, \&
  Sobral}]{Chauke2018}
Chauke, P., van~der Wel, A., Pacifici, C., {et~al.} 2018, \apj, 861, 13,
  \dodoi{10.3847/1538-4357/aac324}

\bibitem[{{Chauke} {et~al.}(2019){Chauke}, {van der Wel}, {Pacifici},
  {Bezanson}, {Wu}, {Gallazzi}, {Straatman}, {Franx}, {Bari{\v{s}}i{\'c}},
  {Bell}, {van Houdt}, {Maseda}, {Muzzin}, {Sobral}, \&
  {Spilker}}]{Chauke2019ApJ}
{Chauke}, P., {van der Wel}, A., {Pacifici}, C., {et~al.} 2019, \apj, 877, 48,
  \dodoi{10.3847/1538-4357/ab164d}

\bibitem[{{Cole} {et~al.}(2005){Cole}, {Percival}, {Peacock}, {Norberg},
  {Baugh}, {Frenk}, {Baldry}, {Bland-Hawthorn}, {Bridges}, {Cannon}, {Colless},
  {Collins}, {Couch}, {Cross}, {Dalton}, {Eke}, {De Propris}, {Driver},
  {Efstathiou}, {Ellis}, {Glazebrook}, {Jackson}, {Jenkins}, {Lahav}, {Lewis},
  {Lumsden}, {Maddox}, {Madgwick}, {Peterson}, {Sutherland}, \&
  {Taylor}}]{Cole2005MNRAS.362..505C}
{Cole}, S., {Percival}, W.~J., {Peacock}, J.~A., {et~al.} 2005, \mnras, 362,
  505, \dodoi{10.1111/j.1365-2966.2005.09318.x}

\bibitem[{{Conroy}(2013)}]{Conroy2013ARAA}
{Conroy}, C. 2013, \araa, 51, 393, \dodoi{10.1146/annurev-astro-082812-141017}

\bibitem[{Cowie {et~al.}(1996)Cowie, Songaila, Hu, \& Cohen}]{Cowie1996}
Cowie, L.~L., Songaila, A., Hu, E.~M., \& Cohen, J.~G. 1996, \aj, 112, 839,
  \dodoi{10.1086/118058}

\bibitem[{Davis(2019)}]{davisExpandingControversy2019}
Davis, T. 2019, Science, 365, 1076, \dodoi{10.1126/science.aay1331}

\bibitem[{{Di Valentino} {et~al.}(2021){Di Valentino}, {Mena}, {Pan},
  {Visinelli}, {Yang}, {Melchiorri}, {Mota}, {Riess}, \&
  {Silk}}]{Valentino2021CQGra}
{Di Valentino}, E., {Mena}, O., {Pan}, S., {et~al.} 2021, Classical and Quantum
  Gravity, 38, 153001, \dodoi{10.1088/1361-6382/ac086d}

\bibitem[{{Eisenstein} {et~al.}(2005){Eisenstein}, {Zehavi}, {Hogg},
  {Scoccimarro}, {Blanton}, {Nichol}, {Scranton}, {Seo}, {Tegmark}, {Zheng},
  {Anderson}, {Annis}, {Bahcall}, {Brinkmann}, {Burles}, {Castander},
  {Connolly}, {Csabai}, {Doi}, {Fukugita}, {Frieman}, {Glazebrook}, {Gunn},
  {Hendry}, {Hennessy}, {Ivezi{\'c}}, {Kent}, {Knapp}, {Lin}, {Loh}, {Lupton},
  {Margon}, {McKay}, {Meiksin}, {Munn}, {Pope}, {Richmond}, {Schlegel},
  {Schneider}, {Shimasaku}, {Stoughton}, {Strauss}, {SubbaRao}, {Szalay},
  {Szapudi}, {Tucker}, {Yanny}, \& {York}}]{Eisenstein2005ApJ...633..560E}
{Eisenstein}, D.~J., {Zehavi}, I., {Hogg}, D.~W., {et~al.} 2005, \apj, 633,
  560, \dodoi{10.1086/466512}

\bibitem[{Estrada-Carpenter {et~al.}(2019)Estrada-Carpenter, Papovich,
  Momcheva, Brammer, Long, Quadri, Bridge, Dickinson, Ferguson, Finkelstein,
  Giavalisco, Gosmeyer, Lotz, Salmon, Skelton, Trump, \&
  Weiner}]{EstradaCarpenter2019}
Estrada-Carpenter, V., Papovich, C., Momcheva, I., {et~al.} 2019, \apj, 870,
  133, \dodoi{10.3847/1538-4357/aaf22e}

\bibitem[{Fanfani(2019)}]{Fanfani2019}
Fanfani, V. 2019, PhD thesis, University of Bologna

\bibitem[{Farr {et~al.}(2019)Farr, Fishbach, Ye, \& Holz}]{Farr2019}
Farr, W.~M., Fishbach, M., Ye, J., \& Holz, D. 2019, ApJL 883 L42 (2019),
  \dodoi{10.3847/2041-8213/ab4284}

\bibitem[{{Ferland} {et~al.}(2017){Ferland}, {Chatzikos}, {Guzm{\'a}n},
  {Lykins}, {van Hoof}, {Williams}, {Abel}, {Badnell}, {Keenan}, {Porter}, \&
  {Stancil}}]{Ferland2017RMxAA}
{Ferland}, G.~J., {Chatzikos}, M., {Guzm{\'a}n}, F., {et~al.} 2017, \rmxaa, 53,
  385.
\newblock \doarXiv{1705.10877}

\bibitem[{Freedman \& Madore(2010)}]{freedmanHubbleConstant2010}
Freedman, W.~L., \& Madore, B.~F. 2010, Annual Review of Astronomy and
  Astrophysics, 48, 673, \dodoi{10.1146/annurev-astro-082708-101829}

\bibitem[{Gallazzi {et~al.}(2014)Gallazzi, Bell, Zibetti, Brinchmann, \&
  Kelson}]{Gallazzi2014}
Gallazzi, A., Bell, E.~F., Zibetti, S., Brinchmann, J., \& Kelson, D.~D. 2014,
  \apj, 788, 72, \dodoi{10.1088/0004-637X/788/1/72}

\bibitem[{{Harris} {et~al.}(2020){Harris}, {Millman}, {van der Walt},
  {Gommers}, {Virtanen}, {Cournapeau}, {Wieser}, {Taylor}, {Berg}, {Smith},
  {Kern}, {Picus}, {Hoyer}, {van Kerkwijk}, {Brett}, {Haldane}, {del R{\'\i}o},
  {Wiebe}, {Peterson}, {G{\'e}rard-Marchant}, {Sheppard}, {Reddy}, {Weckesser},
  {Abbasi}, {Gohlke}, \& {Oliphant}}]{Harris2020}
{Harris}, C.~R., {Millman}, K.~J., {van der Walt}, S.~J., {et~al.} 2020, \nat,
  585, 357, \dodoi{10.1038/s41586-020-2649-2}

\bibitem[{{Holz} \& {Hughes}(2005)}]{Holz2005ApJ}
{Holz}, D.~E., \& {Hughes}, S.~A. 2005, \apj, 629, 15, \dodoi{10.1086/431341}

\bibitem[{Hubble(1936)}]{1936rene.book.....H}
Hubble, E. 1936, The Realm of the Nebulae (New Haven: Yale University Press)

\bibitem[{Hunter(2007)}]{Hunter2007}
Hunter, J.~D. 2007, Computing in Science Engineering, 9, 90,
  \dodoi{10.1109/MCSE.2007.55}

\bibitem[{{Ilbert} {et~al.}(2013){Ilbert}, {McCracken}, {Le F{\`e}vre},
  {Capak}, {Dunlop}, {Karim}, {Renzini}, {Caputi}, \& {Boissier}}]{Ilbert2013}
{Ilbert}, O., {McCracken}, H.~J., {Le F{\`e}vre}, O., {et~al.} 2013, in
  SF2A-2013: Proceedings of the Annual meeting of the French Society of
  Astronomy and Astrophysics, ed. L.~{Cambresy}, F.~{Martins}, E.~{Nuss}, \&
  A.~{Palacios}, 545--548

\bibitem[{{Jimenez} \& {Loeb}(2002)}]{Jimenez2002ApJ}
{Jimenez}, R., \& {Loeb}, A. 2002, \apj, 573, 37, \dodoi{10.1086/340549}

\bibitem[{{Jimenez} {et~al.}(2003){Jimenez}, {Verde}, {Treu}, \&
  {Stern}}]{Jimenez:2003}
{Jimenez}, R., {Verde}, L., {Treu}, T., \& {Stern}, D. 2003, \apj, 593, 622,
  \dodoi{10.1086/376595}

\bibitem[{{Laigle} {et~al.}(2016){Laigle}, {McCracken}, {Ilbert}, {Hsieh},
  {Davidzon}, {Capak}, {Hasinger}, {Silverman}, {Pichon}, {Coupon}, {Aussel},
  {Le Borgne}, {Caputi}, {Cassata}, {Chang}, {Civano}, {Dunlop}, {Fynbo},
  {Kartaltepe}, {Koekemoer}, {Le F{\`e}vre}, {Le Floc'h}, {Leauthaud}, {Lilly},
  {Lin}, {Marchesi}, {Milvang-Jensen}, {Salvato}, {Sanders}, {Scoville},
  {Smolcic}, {Stockmann}, {Taniguchi}, {Tasca}, {Toft}, {Vaccari}, \&
  {Zabl}}]{Laigle2016ApJS}
{Laigle}, C., {McCracken}, H.~J., {Ilbert}, O., {et~al.} 2016, \apjs, 224, 24,
  \dodoi{10.3847/0067-0049/224/2/24}

\bibitem[{{Laureijs} {et~al.}(2011){Laureijs}, {Amiaux}, {Arduini},
  {Augu{\`e}res}, {Brinchmann}, {Cole}, {Cropper}, {Dabin}, {Duvet}, {Ealet},
  {Garilli}, {Gondoin}, {Guzzo}, {Hoar}, {Hoekstra}, \&
  https://www.overleaf.com/4448358379csmrynjykryg}]{Laureijs2011}
{Laureijs}, R., {Amiaux}, J., {Arduini}, S., {et~al.} 2011, arXiv e-prints,
  arXiv:1110.3193.
\newblock \doarXiv{1110.3193}

\bibitem[{{Le F{\`e}vre} {et~al.}(2003){Le F{\`e}vre}, {Saisse}, {Mancini},
  {Brau-Nogue}, {Caputi}, {Castinel}, {D'Odorico}, {Garilli}, {Kissler-Patig},
  {Lucuix}, {Mancini}, {Pauget}, {Sciarretta}, {Scodeggio}, {Tresse}, \&
  {Vettolani}}]{lefevre2003SPIE.4841.1670L}
{Le F{\`e}vre}, O., {Saisse}, M., {Mancini}, D., {et~al.} 2003, in Society of
  Photo-Optical Instrumentation Engineers (SPIE) Conference Series, Vol. 4841,
  Instrument Design and Performance for Optical/Infrared Ground-based
  Telescopes, ed. M.~{Iye} \& A.~F.~M. {Moorwood}, 1670--1681,
  \dodoi{10.1117/12.460959}

\bibitem[{{Leja} {et~al.}(2019){Leja}, {Carnall}, {Johnson}, {Conroy}, \&
  {Speagle}}]{Leja2019ApJ}
{Leja}, J., {Carnall}, A.~C., {Johnson}, B.~D., {Conroy}, C., \& {Speagle},
  J.~S. 2019, \apj, 876, 3, \dodoi{10.3847/1538-4357/ab133c}

\bibitem[{{Ma} \& {Zhang}(2011)}]{Ma2011ApJ}
{Ma}, C., \& {Zhang}, T.-J. 2011, \apj, 730, 74,
  \dodoi{10.1088/0004-637X/730/2/74}

\bibitem[{Moresco(2011)}]{morescoEARLYTYPEGALAXIESPROBES2011}
Moresco, M. 2011, PhD thesis, University of Bologna.
\newblock \url{http://amsdottorato.unibo.it/3725/}

\bibitem[{{Moresco}(2015)}]{Moresco:2015}
{Moresco}, M. 2015, \mnras, 450, L16, \dodoi{10.1093/mnrasl/slv037}

\bibitem[{Moresco {et~al.}(2011)Moresco, Jimenez, Cimatti, \&
  Pozzetti}]{Moresco2011jcap}
Moresco, M., Jimenez, R., Cimatti, A., \& Pozzetti, L. 2011, \jcap, 2011, 045,
  \dodoi{10.1088/1475-7516/2011/03/045}

\bibitem[{{Moresco} {et~al.}(2020){Moresco}, {Jimenez}, {Verde}, {Cimatti}, \&
  {Pozzetti}}]{Moresco2020ApJ}
{Moresco}, M., {Jimenez}, R., {Verde}, L., {Cimatti}, A., \& {Pozzetti}, L.
  2020, \apj, 898, 82, \dodoi{10.3847/1538-4357/ab9eb0}

\bibitem[{{Moresco} {et~al.}(2018){Moresco}, {Jimenez}, {Verde}, {Pozzetti},
  {Cimatti}, \& {Citro}}]{Moresco2018ApJ}
{Moresco}, M., {Jimenez}, R., {Verde}, L., {et~al.} 2018, \apj, 868, 84,
  \dodoi{10.3847/1538-4357/aae829}

\bibitem[{{Moresco} {et~al.}(2012){Moresco}, {Verde}, {Pozzetti}, {Jimenez}, \&
  {Cimatti}}]{Moresco:2012}
{Moresco}, M., {Verde}, L., {Pozzetti}, L., {Jimenez}, R., \& {Cimatti}, A.
  2012, \jcap, 7, 053, \dodoi{10.1088/1475-7516/2012/07/053}

\bibitem[{Moresco {et~al.}(2012)Moresco, Cimatti, Jimenez, Pozzetti, Zamorani,
  Bolzonella, Dunlop, Lamareille, Mignoli, Pearce, Rosati, Stern, Verde, Zucca,
  Carollo, Contini, Kneib, Le~F{\`e}vre, Lilly, Mainieri, Renzini, Scodeggio,
  Balestra, Gobat, McLure, Bardelli, Bongiorno, Caputi, Cucciati, De~La~Torre,
  De~Ravel, Franzetti, Garilli, Iovino, Kampczyk, Knobel, Kova{\v c},
  Le~Borgne, Le~Brun, Maier, Pell{\'o}, Peng, {Perez-Montero}, Presotto,
  Silverman, Tanaka, Tasca, Tresse, Vergani, Almaini, Barnes, Bordoloi,
  Bradshaw, Cappi, Chuter, Cirasuolo, Coppa, Diener, Foucaud, Hartley,
  Kamionkowski, Koekemoer, {L{\'o}pez-Sanjuan}, McCracken, Nair, Oesch,
  Stanford, \& Welikala}]{Moresco2012}
Moresco, M., Cimatti, A., Jimenez, R., {et~al.} 2012, Journal of Cosmology and
  Astroparticle Physics, 2012, \dodoi{10.1088/1475-7516/2012/08/006}

\bibitem[{{Moresco} {et~al.}(2013){Moresco}, {Pozzetti}, {Cimatti}, {Zamorani},
  {Bolzonella}, {Lamareille}, {Mignoli}, {Zucca}, {Lilly}, {Carollo},
  {Contini}, {Kneib}, {Le F{\`e}vre}, {Mainieri}, {Renzini}, {Scodeggio},
  {Bardelli}, {Bongiorno}, {Caputi}, {Cucciati}, {de la Torre}, {de Ravel},
  {Franzetti}, {Garilli}, {Iovino}, {Kampczyk}, {Knobel}, {Kova{\v{c}}}, {Le
  Borgne}, {Le Brun}, {Maier}, {Pell{\'o}}, {Peng}, {Perez-Montero},
  {Presotto}, {Silverman}, {Tanaka}, {Tasca}, {Tresse}, {Vergani}, {Barnes},
  {Bordoloi}, {Cappi}, {Diener}, {Koekemoer}, {Le Floc'h}, {L{\'o}pez-Sanjuan},
  {McCracken}, {Nair}, {Oesch}, {Scarlata}, {Scoville}, \&
  {Welikala}}]{Moresco2013}
{Moresco}, M., {Pozzetti}, L., {Cimatti}, A., {et~al.} 2013, \aap, 558, A61,
  \dodoi{10.1051/0004-6361/201321797}

\bibitem[{{Moresco} {et~al.}(2016){Moresco}, {Pozzetti}, {Cimatti}, {Jimenez},
  {Maraston}, {Verde}, {Thomas}, {Citro}, {Tojeiro}, \&
  {Wilkinson}}]{Moresco:2016}
---. 2016, \jcap, 5, 014, \dodoi{10.1088/1475-7516/2016/05/014}

\bibitem[{{Moresco} {et~al.}(2022){Moresco}, {Amati}, {Amendola}, {Birrer},
  {Blakeslee}, {Cantiello}, {Cimatti}, {Darling}, {Della Valle}, {Fishbach},
  {Grillo}, {Hamaus}, {Holz}, {Izzo}, {Jimenez}, {Lusso}, {Meneghetti},
  {Piedipalumbo}, {Pisani}, {Pourtsidou}, {Pozzetti}, {Quartin}, {Risaliti},
  {Rosati}, \& {Verde}}]{Moresco2022}
{Moresco}, M., {Amati}, L., {Amendola}, L., {et~al.} 2022, arXiv e-prints,
  arXiv:2201.07241.
\newblock \doarXiv{2201.07241}

\bibitem[{{Muzzin} {et~al.}(2013){Muzzin}, {Marchesini}, {Stefanon}, {Franx},
  {Milvang-Jensen}, {Dunlop}, {Fynbo}, {Brammer}, {Labb{\'e}}, \& {van
  Dokkum}}]{Muzzin2013ApJS}
{Muzzin}, A., {Marchesini}, D., {Stefanon}, M., {et~al.} 2013, \apjs, 206, 8,
  \dodoi{10.1088/0067-0049/206/1/8}

\bibitem[{Onodera {et~al.}(2015)Onodera, Carollo, Renzini, Cappellari, Mancini,
  Arimoto, Daddi, Gobat, Strazzullo, Tacchella, \& Yamada}]{Onodera2015}
Onodera, M., Carollo, C.~M., Renzini, A., {et~al.} 2015, \apj, 808, 161,
  \dodoi{10.1088/0004-637X/808/2/161}

\bibitem[{Pacifici {et~al.}(2016)Pacifici, Kassin, Weiner, Holden, Gardner,
  Faber, Ferguson, Koo, Primack, Bell, Dekel, Gawiser, Giavalisco, Rafelski,
  Simons, Barro, Croton, Dav{\'e}, Fontana, Grogin, Koekemoer, Lee, Salmon,
  Somerville, \& Behroozi}]{Pacifici2016}
Pacifici, C., Kassin, S.~A., Weiner, B.~J., {et~al.} 2016, \apj, 832, 79,
  \dodoi{10.3847/0004-637X/832/1/79}

\bibitem[{{Peng} {et~al.}(2010){Peng}, {Lilly}, {Kova{\v{c}}}, {Bolzonella},
  {Pozzetti}, {Renzini}, {Zamorani}, {Ilbert}, {Knobel}, {Iovino}, {Maier},
  {Cucciati}, {Tasca}, {Carollo}, {Silverman}, {Kampczyk}, {de Ravel},
  {Sanders}, {Scoville}, {Contini}, {Mainieri}, {Scodeggio}, {Kneib}, {Le
  F{\`e}vre}, {Bardelli}, {Bongiorno}, {Caputi}, {Coppa}, {de la Torre},
  {Franzetti}, {Garilli}, {Lamareille}, {Le Borgne}, {Le Brun}, {Mignoli},
  {Perez Montero}, {Pello}, {Ricciardelli}, {Tanaka}, {Tresse}, {Vergani},
  {Welikala}, {Zucca}, {Oesch}, {Abbas}, {Barnes}, {Bordoloi}, {Bottini},
  {Cappi}, {Cassata}, {Cimatti}, {Fumana}, {Hasinger}, {Koekemoer},
  {Leauthaud}, {Maccagni}, {Marinoni}, {McCracken}, {Memeo}, {Meneux}, {Nair},
  {Porciani}, {Presotto}, \& {Scaramella}}]{Peng2010ApJ}
{Peng}, Y.-j., {Lilly}, S.~J., {Kova{\v{c}}}, K., {et~al.} 2010, \apj, 721,
  193, \dodoi{10.1088/0004-637X/721/1/193}

\bibitem[{{Percival} {et~al.}(2001){Percival}, {Baugh}, {Bland-Hawthorn},
  {Bridges}, {Cannon}, {Cole}, {Colless}, {Collins}, {Couch}, {Dalton}, {De
  Propris}, {Driver}, {Efstathiou}, {Ellis}, {Frenk}, {Glazebrook}, {Jackson},
  {Lahav}, {Lewis}, {Lumsden}, {Maddox}, {Moody}, {Norberg}, {Peacock},
  {Peterson}, {Sutherland}, \& {Taylor}}]{Percival2001MNRAS.327.1297P}
{Percival}, W.~J., {Baugh}, C.~M., {Bland-Hawthorn}, J., {et~al.} 2001, \mnras,
  327, 1297, \dodoi{10.1046/j.1365-8711.2001.04827.x}

\bibitem[{Perlmutter {et~al.}(1999)Perlmutter, Aldering, Goldhaber, Knop,
  Nugent, Castro, Deustua, Fabbro, Goobar, Groom, Hook, Kim, Kim, Lee, Nunes,
  Pain, Pennypacker, Quimby, Lidman, Ellis, Irwin, McMahon, Ruiz-Lapuente,
  Walton, Schaefer, Boyle, Filippenko, Matheson, Fruchter, Panagia, Newberg,
  Couch, \& Project}]{Perlmutter1999}
Perlmutter, S., Aldering, G., Goldhaber, G., {et~al.} 1999, \apj, 517, 565,
  \dodoi{10.1086/307221}

\bibitem[{{Planck Collaboration} {et~al.}(2014){Planck Collaboration}, {Ade},
  {Aghanim}, {Armitage-Caplan}, {Arnaud}, {Ashdown}, {Atrio-Barandela},
  {Aumont}, {Baccigalupi}, {Banday}, {Barreiro}, {Bartlett}, {Battaner},
  {Benabed}, {Beno{\^\i}t}, {Benoit-L{\'e}vy}, {Bernard}, {Bersanelli},
  {Bielewicz}, {Bobin}, {Bock}, {Bonaldi}, {Bond}, {Borrill}, {Bouchet},
  {Bridges}, {Bucher}, {Burigana}, {Butler}, {Calabrese}, {Cappellini},
  {Cardoso}, {Catalano}, {Challinor}, {Chamballu}, {Chary}, {Chen}, {Chiang},
  {Chiang}, {Christensen}, {Church}, {Clements}, {Colombi}, {Colombo},
  {Couchot}, {Coulais}, {Crill}, {Curto}, {Cuttaia}, {Danese}, {Davies},
  {Davis}, {de Bernardis}, {de Rosa}, {de Zotti}, {Delabrouille}, {Delouis},
  {D{\'e}sert}, {Dickinson}, {Diego}, {Dolag}, {Dole}, {Donzelli}, {Dor{\'e}},
  {Douspis}, {Dunkley}, {Dupac}, {Efstathiou}, {Elsner}, {En{\ss}lin},
  {Eriksen}, {Finelli}, {Forni}, {Frailis}, {Fraisse}, {Franceschi}, {Gaier},
  {Galeotta}, {Galli}, {Ganga}, {Giard}, {Giardino}, {Giraud-H{\'e}raud},
  {Gjerl{\o}w}, {Gonz{\'a}lez-Nuevo}, {G{\'o}rski}, {Gratton}, {Gregorio},
  {Gruppuso}, {Gudmundsson}, {Haissinski}, {Hamann}, {Hansen}, {Hanson},
  {Harrison}, {Henrot-Versill{\'e}}, {Hern{\'a}ndez-Monteagudo}, {Herranz},
  {Hildebrandt}, {Hivon}, {Hobson}, {Holmes}, {Hornstrup}, {Hou}, {Hovest},
  {Huffenberger}, {Jaffe}, {Jaffe}, {Jewell}, {Jones}, {Juvela},
  {Keih{\"a}nen}, {Keskitalo}, {Kisner}, {Kneissl}, {Knoche}, {Knox}, {Kunz},
  {Kurki-Suonio}, {Lagache}, {L{\"a}hteenm{\"a}ki}, {Lamarre}, {Lasenby},
  {Lattanzi}, {Laureijs}, {Lawrence}, {Leach}, {Leahy}, {Leonardi},
  {Le{\'o}n-Tavares}, {Lesgourgues}, {Lewis}, {Liguori}, {Lilje},
  {Linden-V{\o}rnle}, {L{\'o}pez-Caniego}, {Lubin}, {Mac{\'\i}as-P{\'e}rez},
  {Maffei}, {Maino}, {Mandolesi}, {Maris}, {Marshall}, {Martin},
  {Mart{\'\i}nez-Gonz{\'a}lez}, {Masi}, {Massardi}, {Matarrese}, {Matthai},
  {Mazzotta}, {Meinhold}, {Melchiorri}, {Melin}, {Mendes}, {Menegoni},
  {Mennella}, {Migliaccio}, {Millea}, {Mitra}, {Miville-Desch{\^e}nes},
  {Moneti}, {Montier}, {Morgante}, {Mortlock}, {Moss}, {Munshi}, {Murphy},
  {Naselsky}, {Nati}, {Natoli}, {Netterfield}, {N{\o}rgaard-Nielsen},
  {Noviello}, {Novikov}, {Novikov}, {O'Dwyer}, {Osborne}, {Oxborrow}, {Paci},
  {Pagano}, {Pajot}, {Paladini}, {Paoletti}, {Partridge}, {Pasian},
  {Patanchon}, {Pearson}, {Pearson}, {Peiris}, {Perdereau}, {Perotto},
  {Perrotta}, {Pettorino}, {Piacentini}, {Piat}, {Pierpaoli}, {Pietrobon},
  {Plaszczynski}, {Platania}, {Pointecouteau}, {Polenta}, {Ponthieu}, {Popa},
  {Poutanen}, {Pratt}, {Pr{\'e}zeau}, {Prunet}, {Puget}, {Rachen}, {Reach},
  {Rebolo}, {Reinecke}, {Remazeilles}, {Renault}, {Ricciardi}, {Riller},
  {Ristorcelli}, {Rocha}, {Rosset}, {Roudier}, {Rowan-Robinson},
  {Rubi{\~n}o-Mart{\'\i}n}, {Rusholme}, {Sandri}, {Santos}, {Savelainen},
  {Savini}, {Scott}, {Seiffert}, {Shellard}, {Spencer}, {Starck}, {Stolyarov},
  {Stompor}, {Sudiwala}, {Sunyaev}, {Sureau}, {Sutton}, {Suur-Uski}, {Sygnet},
  {Tauber}, {Tavagnacco}, {Terenzi}, {Toffolatti}, {Tomasi}, {Tristram},
  {Tucci}, {Tuovinen}, {T{\"u}rler}, {Umana}, {Valenziano}, {Valiviita}, {Van
  Tent}, {Vielva}, {Villa}, {Vittorio}, {Wade}, {Wandelt}, {Wehus}, {White},
  {White}, {Wilkinson}, {Yvon}, {Zacchei}, \& {Zonca}}]{Planck2014AA}
{Planck Collaboration}, {Ade}, P.~A.~R., {Aghanim}, N., {et~al.} 2014, \aap,
  571, A16, \dodoi{10.1051/0004-6361/201321591}

\bibitem[{{Planck Collaboration} {et~al.}(2020){Planck Collaboration},
  {Aghanim}, {Akrami}, {Ashdown}, {Aumont}, {Baccigalupi}, {Ballardini},
  {Banday}, {Barreiro}, {Bartolo}, {Basak}, {Battye}, {Benabed}, {Bernard},
  {Bersanelli}, {Bielewicz}, {Bock}, {Bond}, {Borrill}, {Bouchet}, {Boulanger},
  {Bucher}, {Burigana}, {Butler}, {Calabrese}, {Cardoso}, {Carron},
  {Challinor}, {Chiang}, {Chluba}, {Colombo}, {Combet}, {Contreras}, {Crill},
  {Cuttaia}, {de Bernardis}, {de Zotti}, {Delabrouille}, {Delouis}, {Di
  Valentino}, {Diego}, {Dor{\'e}}, {Douspis}, {Ducout}, {Dupac}, {Dusini},
  {Efstathiou}, {Elsner}, {En{\ss}lin}, {Eriksen}, {Fantaye}, {Farhang},
  {Fergusson}, {Fernandez-Cobos}, {Finelli}, {Forastieri}, {Frailis},
  {Fraisse}, {Franceschi}, {Frolov}, {Galeotta}, {Galli}, {Ganga},
  {G{\'e}nova-Santos}, {Gerbino}, {Ghosh}, {Gonz{\'a}lez-Nuevo}, {G{\'o}rski},
  {Gratton}, {Gruppuso}, {Gudmundsson}, {Hamann}, {Handley}, {Hansen},
  {Herranz}, {Hildebrandt}, {Hivon}, {Huang}, {Jaffe}, {Jones}, {Karakci},
  {Keih{\"a}nen}, {Keskitalo}, {Kiiveri}, {Kim}, {Kisner}, {Knox},
  {Krachmalnicoff}, {Kunz}, {Kurki-Suonio}, {Lagache}, {Lamarre}, {Lasenby},
  {Lattanzi}, {Lawrence}, {Le Jeune}, {Lemos}, {Lesgourgues}, {Levrier},
  {Lewis}, {Liguori}, {Lilje}, {Lilley}, {Lindholm}, {L{\'o}pez-Caniego},
  {Lubin}, {Ma}, {Mac{\'\i}as-P{\'e}rez}, {Maggio}, {Maino}, {Mandolesi},
  {Mangilli}, {Marcos-Caballero}, {Maris}, {Martin}, {Martinelli},
  {Mart{\'\i}nez-Gonz{\'a}lez}, {Matarrese}, {Mauri}, {McEwen}, {Meinhold},
  {Melchiorri}, {Mennella}, {Migliaccio}, {Millea}, {Mitra},
  {Miville-Desch{\^e}nes}, {Molinari}, {Montier}, {Morgante}, {Moss}, {Natoli},
  {N{\o}rgaard-Nielsen}, {Pagano}, {Paoletti}, {Partridge}, {Patanchon},
  {Peiris}, {Perrotta}, {Pettorino}, {Piacentini}, {Polastri}, {Polenta},
  {Puget}, {Rachen}, {Reinecke}, {Remazeilles}, {Renzi}, {Rocha}, {Rosset},
  {Roudier}, {Rubi{\~n}o-Mart{\'\i}n}, {Ruiz-Granados}, {Salvati}, {Sandri},
  {Savelainen}, {Scott}, {Shellard}, {Sirignano}, {Sirri}, {Spencer},
  {Sunyaev}, {Suur-Uski}, {Tauber}, {Tavagnacco}, {Tenti}, {Toffolatti},
  {Tomasi}, {Trombetti}, {Valenziano}, {Valiviita}, {Van Tent}, {Vibert},
  {Vielva}, {Villa}, {Vittorio}, {Wandelt}, {Wehus}, {White}, {White},
  {Zacchei}, \& {Zonca}}]{Planck2020AA}
{Planck Collaboration}, {Aghanim}, N., {Akrami}, Y., {et~al.} 2020, \aap, 641,
  A6, \dodoi{10.1051/0004-6361/201833910}

\bibitem[{{Pozzetti} {et~al.}(2010{\natexlab{a}}){Pozzetti}, {Bolzonella},
  {Zucca}, {Zamorani}, {Lilly}, {Renzini}, {Moresco}, {Mignoli}, {Cassata},
  {Tasca}, {Lamareille}, {Maier}, {Meneux}, {Halliday}, {Oesch}, {Vergani},
  {Caputi}, {Kova{\v{c}}}, {Cimatti}, {Cucciati}, {Iovino}, {Peng}, {Carollo},
  {Contini}, {Kneib}, {Le F{\'e}vre}, {Mainieri}, {Scodeggio}, {Bardelli},
  {Bongiorno}, {Coppa}, {de la Torre}, {de Ravel}, {Franzetti}, {Garilli},
  {Kampczyk}, {Knobel}, {Le Borgne}, {Le Brun}, {Pell{\`o}}, {Perez Montero},
  {Ricciardelli}, {Silverman}, {Tanaka}, {Tresse}, {Abbas}, {Bottini}, {Cappi},
  {Guzzo}, {Koekemoer}, {Leauthaud}, {Maccagni}, {Marinoni}, {McCracken},
  {Memeo}, {Porciani}, {Scaramella}, {Scarlata}, \& {Scoville}}]{Pozzetti2010}
{Pozzetti}, L., {Bolzonella}, M., {Zucca}, E., {et~al.} 2010{\natexlab{a}},
  \aap, 523, A13, \dodoi{10.1051/0004-6361/200913020}

\bibitem[{{Pozzetti} {et~al.}(2010{\natexlab{b}}){Pozzetti}, {Bolzonella},
  {Zucca}, {Zamorani}, {Lilly}, {Renzini}, {Moresco}, {Mignoli}, {Cassata},
  {Tasca}, {Lamareille}, {Maier}, {Meneux}, {Halliday}, {Oesch}, {Vergani},
  {Caputi}, {Kova{\v{c}}}, {Cimatti}, {Cucciati}, {Iovino}, {Peng}, {Carollo},
  {Contini}, {Kneib}, {Le F{\'e}vre}, {Mainieri}, {Scodeggio}, {Bardelli},
  {Bongiorno}, {Coppa}, {de la Torre}, {de Ravel}, {Franzetti}, {Garilli},
  {Kampczyk}, {Knobel}, {Le Borgne}, {Le Brun}, {Pell{\`o}}, {Perez Montero},
  {Ricciardelli}, {Silverman}, {Tanaka}, {Tresse}, {Abbas}, {Bottini}, {Cappi},
  {Guzzo}, {Koekemoer}, {Leauthaud}, {Maccagni}, {Marinoni}, {McCracken},
  {Memeo}, {Porciani}, {Scaramella}, {Scarlata}, \&
  {Scoville}}]{Pozzetti2010AA}
---. 2010{\natexlab{b}}, \aap, 523, A13, \dodoi{10.1051/0004-6361/200913020}

\bibitem[{{Ratsimbazafy} {et~al.}(2017){Ratsimbazafy}, {Loubser}, {Crawford},
  {Cress}, {Bassett}, {Nichol}, \& {V{\"a}is{\"a}nen}}]{Ratsimbazafy:2017}
{Ratsimbazafy}, A.~L., {Loubser}, S.~I., {Crawford}, S.~M., {et~al.} 2017,
  \mnras, 467, 3239, \dodoi{10.1093/mnras/stx301}

\bibitem[{Riess(2020)}]{Riess2020}
Riess, A.~G. 2020, Nature Reviews Physics, 2, 10,
  \dodoi{10.1038/s42254-019-0137-0}

\bibitem[{Riess {et~al.}(1998)Riess, Filippenko, Challis, Clocchiatti, Diercks,
  Garnavich, Gilliland, Hogan, Jha, Kirshner, Leibundgut, Phillips, Reiss,
  Schmidt, Schommer, Smith, Spyromilio, Stubbs, Suntzeff, \& Tonry}]{Riess1998}
Riess, A.~G., Filippenko, A.~V., Challis, P., {et~al.} 1998, \aj, 116, 1009,
  \dodoi{10.1086/300499}

\bibitem[{{Schutz}(1986)}]{Schutz1986Nature}
{Schutz}, B.~F. 1986, \nat, 323, 310, \dodoi{10.1038/323310a0}

\bibitem[{{Scolnic} {et~al.}(2018){Scolnic}, {Jones}, {Rest}, {Pan},
  {Chornock}, {Foley}, {Huber}, {Kessler}, {Narayan}, {Riess}, {Rodney},
  {Berger}, {Brout}, {Challis}, {Drout}, {Finkbeiner}, {Lunnan}, {Kirshner},
  {Sanders}, {Schlafly}, {Smartt}, {Stubbs}, {Tonry}, {Wood-Vasey}, {Foley},
  {Hand}, {Johnson}, {Burgett}, {Chambers}, {Draper}, {Hodapp}, {Kaiser},
  {Kudritzki}, {Magnier}, {Metcalfe}, {Bresolin}, {Gall}, {Kotak}, {McCrum}, \&
  {Smith}}]{Scolnic2018ApJ}
{Scolnic}, D.~M., {Jones}, D.~O., {Rest}, A., {et~al.} 2018, \apj, 859, 101,
  \dodoi{10.3847/1538-4357/aab9bb}

\bibitem[{{Simon} {et~al.}(2005){Simon}, {Verde}, \& {Jimenez}}]{Simon:2005}
{Simon}, J., {Verde}, L., \& {Jimenez}, R. 2005, \prd, 71, 123001,
  \dodoi{10.1103/PhysRevD.71.123001}

\bibitem[{{Smoot} {et~al.}(1992){Smoot}, {Bennett}, {Kogut}, {Wright}, {Aymon},
  {Boggess}, {Cheng}, {de Amici}, {Gulkis}, {Hauser}, {Hinshaw}, {Jackson},
  {Janssen}, {Kaita}, {Kelsall}, {Keegstra}, {Lineweaver}, {Loewenstein},
  {Lubin}, {Mather}, {Meyer}, {Moseley}, {Murdock}, {Rokke}, {Silverberg},
  {Tenorio}, {Weiss}, \& {Wilkinson}}]{Smoot1992ApJ}
{Smoot}, G.~F., {Bennett}, C.~L., {Kogut}, A., {et~al.} 1992, \apjl, 396, L1,
  \dodoi{10.1086/186504}

\bibitem[{{Stern} {et~al.}(2010){Stern}, {Jimenez}, {Verde}, {Kamionkowski}, \&
  {Stanford}}]{Stern:2010}
{Stern}, D., {Jimenez}, R., {Verde}, L., {Kamionkowski}, M., \& {Stanford},
  S.~A. 2010, \jcap, 2, 008, \dodoi{10.1088/1475-7516/2010/02/008}

\bibitem[{Straatman {et~al.}(2018)Straatman, van~der Wel, Bezanson, Pacifici,
  Gallazzi, Wu, Noeske, Bari{\v s}i{\'c}, Bell, Brammer, Calhau, Chauke, Franx,
  Houdt, Labb{\'e}, Maseda, {Mu{\~n}oz-Mateos}, Muzzin, van~de Sande, Sobral,
  \& Spilker}]{Straatman2018}
Straatman, C. M.~S., van~der Wel, A., Bezanson, R., {et~al.} 2018, The
  Astrophysical Journal Supplement Series, 239, 27,
  \dodoi{10.3847/1538-4365/aae37a}

\bibitem[{{Sullivan} {et~al.}(2011){Sullivan}, {Guy}, {Conley}, {Regnault},
  {Astier}, {Balland}, {Basa}, {Carlberg}, {Fouchez}, {Hardin}, {Hook},
  {Howell}, {Pain}, {Palanque-Delabrouille}, {Perrett}, {Pritchet}, {Rich},
  {Ruhlmann-Kleider}, {Balam}, {Baumont}, {Ellis}, {Fabbro}, {Fakhouri},
  {Fourmanoit}, {Gonz{\'a}lez-Gait{\'a}n}, {Graham}, {Hudson}, {Hsiao},
  {Kronborg}, {Lidman}, {Mourao}, {Neill}, {Perlmutter}, {Ripoche}, {Suzuki},
  \& {Walker}}]{Sullivan2011ApJ}
{Sullivan}, M., {Guy}, J., {Conley}, A., {et~al.} 2011, \apj, 737, 102,
  \dodoi{10.1088/0004-637X/737/2/102}

\bibitem[{{Suzuki} {et~al.}(2012){Suzuki}, {Rubin}, {Lidman}, {Aldering},
  {Amanullah}, {Barbary}, {Barrientos}, {Botyanszki}, {Brodwin}, {Connolly},
  {Dawson}, {Dey}, {Doi}, {Donahue}, {Deustua}, {Eisenhardt}, {Ellingson},
  {Faccioli}, {Fadeyev}, {Fakhouri}, {Fruchter}, {Gilbank}, {Gladders},
  {Goldhaber}, {Gonzalez}, {Goobar}, {Gude}, {Hattori}, {Hoekstra}, {Hsiao},
  {Huang}, {Ihara}, {Jee}, {Johnston}, {Kashikawa}, {Koester}, {Konishi},
  {Kowalski}, {Linder}, {Lubin}, {Melbourne}, {Meyers}, {Morokuma}, {Munshi},
  {Mullis}, {Oda}, {Panagia}, {Perlmutter}, {Postman}, {Pritchard}, {Rhodes},
  {Ripoche}, {Rosati}, {Schlegel}, {Spadafora}, {Stanford}, {Stanishev},
  {Stern}, {Strovink}, {Takanashi}, {Tokita}, {Wagner}, {Wang}, {Yasuda},
  {Yee}, \& {Supernova Cosmology Project}}]{Suzuki2012ApJ}
{Suzuki}, N., {Rubin}, D., {Lidman}, C., {et~al.} 2012, \apj, 746, 85,
  \dodoi{10.1088/0004-637X/746/1/85}

\bibitem[{{Swetz} {et~al.}(2011){Swetz}, {Ade}, {Amiri}, {Appel},
  {Battistelli}, {Burger}, {Chervenak}, {Devlin}, {Dicker}, {Doriese},
  {D{\"u}nner}, {Essinger-Hileman}, {Fisher}, {Fowler}, {Halpern},
  {Hasselfield}, {Hilton}, {Hincks}, {Irwin}, {Jarosik}, {Kaul}, {Klein},
  {Lau}, {Limon}, {Marriage}, {Marsden}, {Martocci}, {Mauskopf}, {Moseley},
  {Netterfield}, {Niemack}, {Nolta}, {Page}, {Parker}, {Staggs}, {Stryzak},
  {Switzer}, {Thornton}, {Tucker}, {Wollack}, \& {Zhao}}]{Swetz2011ApJS}
{Swetz}, D.~S., {Ade}, P.~A.~R., {Amiri}, M., {et~al.} 2011, \apjs, 194, 41,
  \dodoi{10.1088/0067-0049/194/2/41}

\bibitem[{Tacchella {et~al.}(2022)Tacchella, Conroy, Faber, Johnson, Leja,
  Barro, Cunningham, Deason, Guhathakurta, Guo, Hernquist, Koo, McKinnon,
  Rockosi, Speagle, van Dokkum, \& Yesuf}]{Tacchella2022}
Tacchella, S., Conroy, C., Faber, S.~M., {et~al.} 2022, \apj, 926, 134,
  \dodoi{10.3847/1538-4357/ac449b}

\bibitem[{Thomas {et~al.}(2011)Thomas, Maraston, \& Johansson}]{Thomas2011}
Thomas, D., Maraston, C., \& Johansson, J. 2011, \mnras, 412, 2183,
  \dodoi{10.1111/j.1365-2966.2010.18049.x}

\bibitem[{{van Dokkum} {et~al.}(2000){van Dokkum}, {Franx}, {Fabricant},
  {Illingworth}, \& {Kelson}}]{Dokkum2000ApJ}
{van Dokkum}, P.~G., {Franx}, M., {Fabricant}, D., {Illingworth}, G.~D., \&
  {Kelson}, D.~D. 2000, \apj, 541, 95, \dodoi{10.1086/309402}

\bibitem[{Verde {et~al.}(2019)Verde, Treu, \& Riess}]{Verde2019}
Verde, L., Treu, T., \& Riess, A.~G. 2019, Nature Astronomy, 3, 891,
  \dodoi{10.1038/s41550-019-0902-0}

\bibitem[{Virtanen {et~al.}(2020)Virtanen, Gommers, Oliphant, Haberland, Reddy,
  Cournapeau, Burovski, Peterson, Weckesser, Bright, {van der Walt}, Brett,
  Wilson, Millman, Mayorov, Nelson, Jones, Kern, Larson, Carey, Polat, Feng,
  Moore, {VanderPlas}, Laxalde, Perktold, Cimrman, Henriksen, Quintero, Harris,
  Archibald, Ribeiro, Pedregosa, {van Mulbregt}, \& {SciPy 1.0
  Contributors}}]{2020SciPy-NMeth}
Virtanen, P., Gommers, R., Oliphant, T.~E., {et~al.} 2020, Nature Methods, 17,
  261, \dodoi{10.1038/s41592-019-0686-2}

\bibitem[{{Wang} {et~al.}(2019){Wang}, {Dickinson}, {Hillenbrand}, {Robberto},
  {Armus}, {Ballardini}, {Barkhouser}, {Bartlett}, {Behroozi}, {Benjamin},
  {Brinchmann}, {Chary}, {Chuang}, {Cimatti}, {Conroy}, {Content}, {Daddi},
  {Donahue}, {Dore}, {Eisenhardt}, {Ferguson}, {Faisst}, {Fraser},
  {Glazebrook}, {Gorjian}, {Helou}, {Hirata}, {Hudson}, {Kirkpatrick},
  {Malhotra}, {Mei}, {Moscardini}, {Newman}, {Ninkov}, {Orsi}, {Ressler},
  {Rhoads}, {Rhodes}, {Ryan}, {Samushia}, {Scarlata}, {Scolnic}, {Seiffert},
  {Shapley}, {Smee}, {Valentino}, {Vorobiev}, \& {Wechsler}}]{Atlasmission}
{Wang}, Y., {Dickinson}, M., {Hillenbrand}, L., {et~al.} 2019, arXiv e-prints,
  arXiv:1909.00070.
\newblock \doarXiv{1909.00070}

\bibitem[{{Weaver} {et~al.}(2022){Weaver}, {Kauffmann}, {Ilbert}, {McCracken},
  {Moneti}, {Toft}, {Brammer}, {Shuntov}, {Davidzon}, {Hsieh}, {Laigle},
  {Anastasiou}, {Jespersen}, {Vinther}, {Capak}, {Casey}, {McPartland},
  {Milvang-Jensen}, {Mobasher}, {Sanders}, {Zalesky}, {Arnouts}, {Aussel},
  {Dunlop}, {Faisst}, {Franx}, {Furtak}, {Fynbo}, {Gould}, {Greve}, {Gwyn},
  {Kartaltepe}, {Kashino}, {Koekemoer}, {Kokorev}, {Le F{\`e}vre}, {Lilly},
  {Masters}, {Magdis}, {Mehta}, {Peng}, {Riechers}, {Salvato}, {Sawicki},
  {Scarlata}, {Scoville}, {Shirley}, {Silverman}, {Sneppen}, {Smolc̆i{\'c}},
  {Steinhardt}, {Stern}, {Tanaka}, {Taniguchi}, {Teplitz}, {Vaccari}, {Wang},
  \& {Zamorani}}]{Weaver2022ApJS}
{Weaver}, J.~R., {Kauffmann}, O.~B., {Ilbert}, O., {et~al.} 2022, \apjs, 258,
  11, \dodoi{10.3847/1538-4365/ac3078}

\bibitem[{{Williams} {et~al.}(2009){Williams}, {Quadri}, {Franx}, {van Dokkum},
  \& {Labb{\'e}}}]{Williams2009ApJ}
{Williams}, R.~J., {Quadri}, R.~F., {Franx}, M., {van Dokkum}, P., \&
  {Labb{\'e}}, I. 2009, \apj, 691, 1879, \dodoi{10.1088/0004-637X/691/2/1879}

\bibitem[{Wu {et~al.}(2020)Wu, Yu, \& Wang}]{wuNewMethodMeasure2020}
Wu, Q., Yu, H., \& Wang, F.~Y. 2020, The Astrophysical Journal, 895, 33,
  \dodoi{10.3847/1538-4357/ab88d2}

\bibitem[{{Zhang} {et~al.}(2014){Zhang}, {Zhang}, {Yuan}, {Liu}, {Zhang}, \&
  {Sun}}]{Zhang:2014}
{Zhang}, C., {Zhang}, H., {Yuan}, S., {et~al.} 2014, \raa, 14, 1221,
  \dodoi{10.1088/1674-4527/14/10/002}

\bibitem[{{Zhang} {et~al.}(2010){Zhang}, {Ma}, \& {Lan}}]{Zhang2010AdAst}
{Zhang}, T.-J., {Ma}, C., \& {Lan}, T. 2010, Advances in Astronomy, 2010,
  184284, \dodoi{10.1155/2010/184284}

\end{thebibliography}
\bibliographystyle{aasjournal}
\end{document}